%&TeX
% version 02/09/06
\input harvmac.tex
%%%%%%%%%%%%%%%%%%%%%%%%%%%
%DEFINITIONS IVAN
\def\Re{{\rm Re ~}}

\def\p{\partial}

%%%%%%%%%GREEK LETTERS%%%%%%%%%%%%%%%%
\def\a{\alpha}

\def\g{\gamma}
\def\d{\delta}

\def\k{\kappa}
\def\l{\lambda}
\def\m{\mu}

\def\s{\sigma}
\def\t{\tau}

\def\z{\zeta }
\def\vp {\varphi}
\def\G{\Gamma}

\def\O{\Omega}
\def\oo{\hat \omega   }

   \def\Re{{\rm Re} }

\def\[{\left[}
\def\]{\right]}
\def\({\left(}
\def\){\right)}
\def\<{\left\langle\,}
\def\>{\,\right\rangle}
\def\hf{ {\frac{1}{2}}}
\def\half{ {1\over 2} }
%%%%%%%%%%%%%%%%%%%%%%%%
\def\inv{^{-1}}

\def\frac#1#2{ {{\textstyle{#1\over#2}}}}
\def\inv{^{\raise.15ex\hbox{${\scriptscriptstyle -}$}\kern-.05em 1}}
\def\hf{ \frac{1}{2} }
  
\def\rb{ \noindent $\bullet$\ \ }
%DEFINITIONS  VALYA

 \def\ra{\rangle}

\def\IR{{ \Bbb R} }
\def\IZ{{ \Bbb Z} }

\def\IC{{ \Bbb C} }
\def\IX{{ \Bbb X} }
\def\IP{{ \Bbb P} }
\def\dC{C\kern-6.5pt I}

\def\CA{{\cal A}}              \def\CC{{\cal C}}
\def\CD{{\cal D}}              
\def\CG{{\cal G}}              
              
              \def\CO{{\cal O}}

       \def\CZ{{\cal Z}}

\def\L{\Lambda}

%%%%%%%%%%%%%%%%%%%

    \def\oo{ {^{_\circ}_{^\circ}}}
\def\tw{ {_{\rm tw}}}
 \def\oi{  0_{_{\infty} } \,}

%%%%%%%Russian fonts%%%%%%%5
\chardef\tempcat=\the\catcode`\@ \catcode`\@=11
\def\cyracc{\def\u##1{\if \i##1\accent"24 i%
    \else \accent"24 ##1\fi }}
\newfam\cyrfam

%%%%% REFS %%%%%%%%%%
\def\hepth#1{{e-Print Archive:  hep-th/}#1}
\def\Sx{S_x  }

\def\gs{ g_{\rm s}}

\def\pt{\p_{\tau}}

  \def\Ltot{ L_{\rm tot}}

\def\rm{ {_{  \langle -\rangle}} }

\def\SRSOS{SRSOS }
\hfuzz 15pt
%\draftmode
\input amssym.def
\input amssym.tex
\input
epsf\def\tfig#1{{
\xdef#1{Fig.\thinspace\the\figno}}Fig.\thinspace\the\figno
\global\advance\figno by1}
%%%%%%%%%%%%%%%%%%DEFINITIONS%%%%%

\input epsf
% font definitions

\def\oo{ {^{_\circ}_{^\circ}}}
\def\tw{ {_{\rm tw}}}

\def\Ga{ \Gamma }

 \def\lll{ {  \rangle \!- }   }
\def\rrr{ {  -\!\langle }}

%%%%%%%%REFERENCES                          %%%%%%%%%%%%%%%%%%%%%%%%%%
%
\lref\DO{ H. Dorn, H.J. Otto: Two and three point functions in
Liouville theory, {\it Nucl.  Phys.} {\bf B 429} (1994) 375,
hep-th/9403141.  }
\lref\ZZtp{A.B. Zamolodchikov and Al.B. Zamolodchikov, Structure
constants and conformal bootstrap in Liouville field theory, {\it
Nucl.  Phys.} {\bf B 477} (1996) 577, hep-th/9506136.  }
\lref\Ta{J. Teschner, On the Liouville three-point function.  {\it
Phys.Lett.}, {\bf B 363} (1995) 65, hep-th/9507109.  }
\lref\PTa{B.~Ponsot, J.~Teschner, ``Liouville bootstrap via harmonic
analysis on a noncompact quantum group'', hep-th/99111110;
``Clebsch-Gordan and Racah-Wigner coefficients for a continuous series
of representations of $U_q(sl(2,\IR))$'', {\it Comm.  Math.  Phys.}
{\bf 224}, 3 (2001), math-qa/0007097.}
    \lref\FZZb{V.~Fateev, A.~B.~Zamolodchikov and A.~B.~Zamolodchikov,
    ``Boundary Liouville field theory.  I: Boundary state and boundary
    two-point function'', hep-th/0001012.  }
\lref\ZZPseudo{ A.~B. Zamolodchikov and A.~B. Zamolodchikov,
``Liouville field theory on a pseudosphere, \hepth{0101152}.  }
\lref\PTtwo{B.~Ponsot, J.~Teschner, ``Boundary Liouville field theory:
Boundary three point function'', Nucl.~Phys.~{\bf B 622} (2002) 309,
\hepth{0110244}.  }
\lref\hosomichi{K.~Hosomichi, "Bulk-Boundary Propagator in Liouville
Theory on a Disc", JHEP {\bf 0111} 044 (2001), \hepth{0108093}.  }
  \lref\GinM{ P. Ginsparg and G. Moore, ``Lectures on 2D gravity and
  2D string theory (TASI 1992)", hep-th/9304011.  }
\lref\DiFrancescoGinsparg{P.~Di Francesco, P.~Ginsparg and
J.~Zinn-Justin,``2-D Gravity and random matrices'', {\it Phys.\
Rept.}\ {\bf 254} (1995) 1, hep-th/9306153.  }
\lref\polchinski{J. Polchinski, ``What is string theory'', {\it
Lectures presented at the 1994 Les Houches Summer School ``Fluctuating
Geometries in Statistical Mechanics and Field Theory''},
\hepth{9411028}.}

\lref\KPZ{ V.~Knizhnik, A.~Polyakov and A.~Zamolodchikov, {\it Mod.
Phys.  Lett.} {\bf A3} 819 (1988); F.~David, {\it Mod.  Phys.  Lett.}
{\bf A3} 1651 (1988); J.~Distler and H.~Kawai, {\it Nucl.  Phys.} {\bf
B321} 509 (1989).  }

\lref\David{ F.~David, ``Loop Equations And Nonperturbative Effects In
Two-Dimensional Quantum Gravity,'' Mod.\ Phys.\ Lett.\ A {\bf 5}, 1019
(1990).  }

 \lref\AlZG{ Al.~Zamolodchikov, arXiv:hep-th/0505063.  }
\lref\KlebanovMQM{ I.~Klebanov, ``String theory in two-dimensions'',
hep-th/9108019.  }
\lref\JevickiQN{ A.~Jevicki, ``Developments in 2-d string theory'',
hep-th/9309115.  }
  \lref\Witten{ E.~Witten, ``Ground ring of two-dimensional string
  theory'', {Nucl.\ Phys.}\ {\bf B 373}, 187 (1992), hep-th/9108004.
  }
\lref\KlebPol{I.R.Klebanov and A.M. Polyakov, ``Interaction of
discrete states in two-dimensional string theory'', {\it Mod.  Phys.
Lett.} {\bf 6} (1991) 3273, \hepth{9109032}.}
\lref\KMS{ D. Kutasov, E. Martinec, N. Seiberg, ``Ground rings and
their modules in 2-D gravity with $c\le1 $ matter", {\it Phys.  Lett.}
{\bf B 276} (1992) 437, \hepth{9111048}.  }
\lref\kachru{S. Kachru, ``Quantum rings and recursion relations in 2D
quantum gravity'', {\it Mod.  Phys.  Lett.} {\bf A7} (1992) 1419,
hep-th/9201072.}
\lref\bershkut{ M. Bershadsky and D. Kutasov, ``Scattering of open and
closed strings in (1+1)-dimensions", {\it Nucl.  Phys.} {\bf B 382}
(1992) 213, \hepth{9204049}.  }
\lref\Gri{M.T. Grisaru, A. Lerda, S. Penati and D. Zanon, {\it Phys.
Lett.} {\bf B234} (1990) 88; {\it Nucl.  Phys.} {\bf B342} (1990)
564.}

\lref\newhat{ I.~R.~Klebanov, J.~Maldacena and N.~Seiberg, ``D-brane
decay in two-dimensional string theory,'' JHEP {\bf 0307}, 045 (2003)
\hepth{0305159}.  }
\lref\SeibergS{ N. Seiberg and D. Shih, ``Branes, rings and matrix
models on minimal (super)string theory", JHEP {\bf 0402} (2004) 021,
hep-th/0312170.  }

%\GaberdielSZ
\lref\GKR{ M.~R.~Gaberdiel, A.~O.~Klemm and I.~Runkel, ``Matrix model
eigenvalue integrals and twist fields in the su(2)-WZW model,'' JHEP
{\bf 0510}, 107 (2005) [arXiv:hep-th/0509040].  }

\lref\HM{T. Hollowood, P. Mansfield, {\bf Phys.  Lett.} {\bf 226B}
(1989) 73.}

\lref\FS{P. Fendley, H. Saleur, {\it Nucl.  Phys.} {\bf B388} (1992)
609.} \lref\FSZ{ P.~Fendley, H.~Saleur and Al.~Zamolodchikov, {\it
Int.J.Mod.Phys.} {\bf A8} 5751 (1993); {\it Int.J.Mod.Phys.} {\bf A8}
5717 (1993).  }

  \lref\BNh{ H. W. Bloete and B. Nienhuis, {\it Phys.  Rev.  Lett}
  {\bf 72} 1372 (1994).  }
\lref\Morozov{ A.~Morozov, {``Integrability and matrix models''},
Phys.\ Usp.\ {\bf 37} (1994) 1 [arXiv:hep-th/9303139].  }
%
%\diFrancescoQF
\lref\diFSZ{ P.~di Francesco, H.~Saleur and J.~B.~Zuber, ``Relations
Between The Coulomb Gas Picture And Conformal Invariance Of
Two-Dimensional Critical Models,'' Journ.  Stat.  Phys.  {\bf 49} 57
(1987).  }
%
%\KostovCY
\lref\KostovCY{ I.~K.~Kostov, ``Boundary ground ring in 2D string
theory'', {\it Nucl.\ Phys.}\ {\bf B 689}, 3 (2004), hep-th/0312301.
}
\lref\BuKa{ D.~V.~Boulatov and V.~A.~Kazakov, ``The Ising Model On
Random Planar Lattice: The Structure Of Phase Transition And The Exact
Critical Exponents,'' {\it Phys.\ Lett.}\ {\bf 186 B}, 379 (1987).  }
\lref\IOn{ I.~K.~Kostov, ``O(N) Vector Model On A Planar Random
Lattice: Spectrum of Anomalous Dimensions'', {\it Mod.\ Phys.\ Lett.}\
{\bf A 4}, 217 (1989).  }

\lref\KostovCG{ I.~K.~Kostov, ``Strings with discrete target space'',
{\it Nucl.\ Phys.}\ {\bf B 376}, 539 (1992), hep-th/9112059.  }
\lref\HiguchiPV{ S.~Higuchi and I.~K.~Kostov, ``Feynman rules for
string field theories with discrete target space'', {\it Phys.\
Lett.}\ {\bf B 357}, 62 (1995), hep-th/9506022.  }
\lref\AlZ{Al.B. Zamolodchikov, The Three-point Function in the Minimal
Liouville Gravity, {Theor.\ Math.\ Phys.}\ {\bf 142}, 183 (2005); On
the Three-point Function in Minimal Liouville Gravity,
\hepth{0505063}.  }
\lref\DF{Vl.S. Dotsenko and V.A. Fateev, ``Four point correlation
functions and the operator algebra in the two-dimensional conformal
invariant theories with the central charge $c < 1$'', {\it Nucl.
Phys.} {\bf B 251} [FS13], 691 (1985).}
  \lref\DiK{ P. Di Francesco, D. Kutasov, ``World sheet and space time
  physics in two dimensional (super) string sheory", {\it Nucl.Phys.}
  {\bf B 375} (1992) 119, hep-th/9109005.  }
\lref\Ion{ I.~K.~Kostov, ``O(N) Vector Model On A Planar Random
Lattice: Spectrum Of Anomalous Dimensions,'' Mod.\ Phys.\ Lett.\ A
{\bf 4}, 217 (1989).  }
\lref\GauK{ M.~Gaudin and I.~Kostov, ``O(N) Model On A Fluctuating
Planar Lattice: Some Exact Results,'' Phys.\ Lett.\ B {\bf 220}, 200
(1989).  } \lref\KPlet{ I.~K.~Kostov and V.~B.~Petkova, Bulk
correlation functions in 2D quantum gravity, hep-th/0505078, to appear
in the Proceedings of the Workshop "Classical and Quantum Integrable
Systems", Dubna, Jan 24-28, 2005.  }
  \lref\BZ{V. Pokrovsky, A. Belavin and Al.  Zamolodchikov, Moduli
  integrals, ground ring and four-point function in minimal Liouville
  gravity, in {\it Polyakov's string: Twenty five years after},
  hep-th/0510214.}
\lref\BM{ A.~Basu and E.~J.~Martinec, ``Boundary ground ring in
minimal string theory,'' {\it Phys.\ Rev.}\ {\bf D} {\bf 72}, 106007
(2005) \hepth{0509142}.}
\lref\BZJ{ A.~A.~Belavin and A.~B.~Zamolodchikov, ``Moduli integrals
and ground ring in minimal Liouville gravity,'' JETP Lett.\ {\bf 82},
7 (2005) [Pisma Zh.\ Eksp.\ Teor.\ Fiz.\ {\bf 82}, 8 (2005)].  }
\lref\DavidHJ{ F.~David, ``Conformal field theories coupled to 2-D
gravity in the conformal gauge'', {\it Mod.\ Phys.\ Lett.}\ {\bf A 3},
1651 (1988).  }
\lref\DistlerJT{ J.~Distler and H.~Kawai, ``Conformal field theory and
2-D quantum gravity or who's afraid of Joseph Liouville?'', {\it
Nucl.\ Phys.}\ {\bf B 321}, 509 (1989).  }
\lref\Tes{J. Teschner, Remarks on Liouville theory with boundary,
hep-th/0009138.}
\lref\SeibergEB{ N.~Seiberg, ``Notes on quantum Liouville theory and
quantum gravity'', {\it Prog.\ Theor.\ Phys.\ Suppl.}\ {\bf 102}, 319
(1990).  }
  \lref\LianGK{ B.~H.~Lian and G.~J.~Zuckerman, ``New selection rules
  and physical states in 2-D gravity: Conformal gauge'', {\it Phys.\
  Lett.}\ {\bf B 254}, 417 (1991).  }
\lref\DFt{Vl.  Dotsenko and V. Fateev, Operator Algebra of
Two-Dimensional Conformal Theories with Central Charge $C\le 1$, {\it
Phys.  Lett.} {\bf 154B}, (1085) 291.}
\lref\EY{ T. Eguchi and S.-K. Yang, ``Deformations of conformal field
theories and soliton equations", {\it Phys.  Lett.  } {\bf B 224} 373
(1989).  }
\lref\FGP{P. Furlan, A.Ch.  Ganchev and V.B. Petkova, ``Remarks on the
quantum group structure of the rational $c<1$ conformal theories'',
{\it Int.  J. Math.  Phys.} {\bf A6} (1991) 4859.}
\lref\ST{A. Strominger and T. Takayanagi, ``Correlators in time like
bulk Liouville theory'', {\it Adv, Theor.  Math.  Phys.} {\bf 7}, 369
(2003), hep-th/0303221.  }
\lref\Sch{V. Schomerus, ``Rolling tachyons from Liouville theory'',
JHEP {\bf 0311} (2003) 043, hep-th/0306026.  }
\lref\IMM{ C. Imbimbo, S. Mahpatra and S. Mukhi, Construction of
Physical states of non-trivial ghost number in $c<1$ string theory,
{\it Nucl.  Phys.} {\bf B 375} (1992) 399.  }
\lref\PZb{V.B. Petkova and J-B. Zuber, The many faces of Ocneanu
cells, {\it Nucl.  Phys.  } {\bf B 603} (2001) 449, hep-th/0101151.  }
\lref\KKOPNS{ D. Kutasov, K. Okuyama, J. Park, N. Seiberg, D. Shih,
``Annulus Amplitudes and ZZ Branes in Minimal String Theory'', JHEP
{\bf 0408} (2004) 026, hep-th/0406030.  }
\lref\PasquierJC{ V.~Pasquier, ``Two-dimensional critical systems
labelled by Dynkin diagrams'', Nucl.\ Phys.\ {\bf B 285}, 162 (1987).
}
\lref\Dijk{ R.~Dijkgraaf, ``Intersection theory, integrable
hierarchies and topological field theory,'' arXiv:hep-th/9201003.  }
\lref\Bulkcft{ I. Kostov and V. Petkova, Non-Rational 2D Quantum
Gravity I. World sheet CFT, hep-th/0512346.  }
% %
\lref\FB{ P.~J.~Forrester and R.~J.~Baxter, {\it J.~Stat.\ Phys.}\
{\bf38} (1985) 435.  }
\lref\DJKMO{ E.~Date, M.~Jimbo, A.~Kuniba, T.~Miwa, and M.~Okado, {\it
Nucl.\ Phys.}\ {\bf B290} [FS20] (1987) 231; {\it Adv.\ Stud.\ Pure
Math.}\ {\bf 16} (1988) 17.  }
\lref\ABF{ G.~E.~Andrews, R.~J.~Baxter, and P.~J.~Forrester, {\it J.\
Stat.\ Phys.}\ {\bf 35} (1984) 193.  }
\lref\Pasq{V.~Pasquier, ``Operator content of the ADE lattice
models'', {\it J. Phys.} {\bf A 20}, 5707 (1987).  }
\lref\Huse{ D.~A.~Huse, {\it Phys.\ Rev.}\ {\bf B 30} (1984) 3908.  }
% %
\lref\Baxter{ R.~J.~Baxter, {\it Ann.\ Phys.}\ {\bf 76} (1973) 1; 26;
48 }
  \lref\nienhuis{B. Nienhuis, {\it J. Stat.  Phys.  } {\bf 34}, 731
  (1984); \ ``Coulomb gas formulation of 2-d phase transitions", {\it
  in} Phase transitions and critical phenomena, Vol.  11, ed.  C.C.
  Domb and J.L. Lebowitz (Academic Press, New York, 1987), ch.1.  }
\lref\KPS{ I.~K.~Kostov, B.~Ponsot and D.~Serban, ``Boundary Liouville
theory and 2D quantum gravity'', {Nucl.\ Phys.}\ {\bf B 683}, 309
(2004), hep-th/0307189.  }
\lref\FodaIN{ O.~Foda and B.~Nienhuis, ``The Coulomb gas
representation of critical RSOS models on the sphere and the torus'',
Nucl.\ Phys.\ {\bf B 324}, 643 (1989).}
  \lref\Idis{ I.~K.~Kostov, ``Strings with discrete target space'',
  {Nucl.\ Phys.}\ {\bf B 376}, 539 (1992), hep-th/9112059.  }
 \lref\KSone{ I.~K.~Kostov and M. Staudacher, ``Multicritical points
 of the $O(n)$ Model on a Random Lattice", {Nucl.\ Phys.}\ {\bf B 384}
 (1992) 459-483.  }
\lref\Onth{ I.~K.~Kostov, ``Thermal flow in the gravitational O(n)
model,'' arXiv:hep-th/0602075.  }
\lref\KostovPN{ I.~K.~Kostov and M.~Staudacher, ``Multicritical phases
of the O(n) model on a random lattice,'' Nucl.\ Phys.\ B {\bf 384},
459 (1992) [arXiv:hep-th/9203030].  }
\lref\KazakovBC{ V.~A.~Kazakov, ``The appearance of matter fields from
quantum fluctuations of 2-D gravity,'' Mod.\ Phys.\ Lett.\ A {\bf 4},
2125 (1989).  }
   \lref\adem{ I. Kostov, ``Gauge Invariant Matrix Model for the
   \^A-\^D-\^E Closed Strings", {\it Phys.  Lett.} {\bf B 297} (1992)
   74-81, hep-th/9208053.  }
\lref\BIPZ{ E.~Brezin, C.~Itzykson, G.~Parisi and J.~B.~Zuber,
``Planar Diagrams,'' {\it Commun.\ Math.\ Phys.}\ {\bf 59}, 35 (1978).
}
\lref\ADEold{ I.K. Kostov, ``The $ADE$ face models on a fluctuating
planar lattice", {\it Nucl.\ Phys.} {\bf B 326}, 583 (1989).  }
\lref\EynardNV{ B.~Eynard and C.~Kristjansen, ``Exact solution of the
O(n) model on a random lattice,'' {\it Nucl.\ Phys.} {\bf B 455}, 577
(1995), hep-th/9506193.  }

  \lref\GinM{ P. Ginsparg and G. Moore, ``Lectures on 2D gravity and
  2D string theory (TASI 1992)", hep-th/9304011.  }
\lref\Icar{ I.~K.~Kostov, ``Solvable statistical models on a random
lattice'', Nucl.\ Phys.\ Proc.\ Suppl.\ {\bf 45A}, 13 (1996),
hep-th/9509124.  } \lref\MMM{ A.~Marshakov, A.~Mironov, A.~Morozov, {
``Generalized matrix models as conformal field theories: Discrete
case''}, Phys.\ Lett.\ B {\bf 265} (1991) 99.  }
  \lref\CMM{S. Kharchev, A. Marshakov, A. Mironov, A. Morozov, S.
  Pakuliak, ``Conformal Matrix Models as an Alternative to
  Conventional Multi-Matrix Models'', {\it Nucl.  Phys.} {\bf B 404}
  (1993) 717, hep-th/9208044.  }
\lref\Higkos{ S.~Higuchi and I.~K.~Kostov, ``Feynman rules for string
field theories with discrete target space,'' Phys.\ Lett.\ B {\bf
357}, 62 (1995), hep-th/9506022.  }
\lref\Iconfm{ I. Kostov, ``Conformal Field Theory Techniques in Random
Matrix models", hep-th/9907060.}
\lref\IHouches{I. Kostov, ``Matrix Models as Conformal Field
Theories'', Lectures given at the summer school ``Applications of
random matrices in physics", Les Houches, June 2004.  }
\lref\SeibergS{ N.~Seiberg and D.~Shih, ``Branes, rings and matrix
models in minimal (super)string theory,'' JHEP {\bf 0402}, 021 (2004),
hep-th/0312170.  }
\lref\MooreIR{ G.~W.~Moore, N.~Seiberg and M.~Staudacher, ``From loops
to states in 2-D quantum gravity,'' {\it Nucl.\ Phys.}\ {\bf B 362},
665 (1991).  }
\lref\KostovHN{ I.~K.~Kostov, ``Multiloop correlators for closed
strings with discrete target space,'' {\it Phys.\ Lett.}\ B {\bf 266},
42 (1991).}
\lref\PZ{V.B. Petkova and J-B. Zuber, ``On structure constants of
$sl(2)$ theories'', {\it Nucl.  Phys.} {\bf B 438}, 347 (1995),
hep-th/9410209.  }
\lref\BPPZ{R. Behrend, P. Pearce, V.B. Petkova and J-B. Zuber, ``On
the classification of bulk and boundary conformal field theories'',
{\it Phys.  Lett.} {\bf B 444}, 163 (1998), hep-th/9809097; ``Boundary
conditions in rational conformal field theories'', {\it Nucl.  Phys.
} {\bf B 579} (2000) 707, hep-th/9908036.  }
\lref\DVV{ R.~Dijkgraaf, H.~L.~Verlinde and E.~P.~Verlinde, ``Loop
Equations And Virasoro Constraints In Nonperturbative 2-D Quantum
Gravity,'' {\it Nucl.\ Phys.}\ B {\bf 348}, 435 (1991);
%
%\lref\FKN{
M.~Fukuma, H.~Kawai and R.~Nakayama, ``Continuum Schwinger-Dyson
Equations And Universal Structures In Two-Dimensional Quantum
Gravity,'' {\it Int.\ J.\ Mod.\ Phys.}\ A {\bf 6}, 1385 (1991).  }
%
%\FukumaNM
\lref\Fukuma{ M.~Fukuma, H.~Irie and S.~Seki, ``Comments on the
D-instanton calculus in (p,p+1) minimal string theory,'' Nucl.\ Phys.\
B {\bf 728}, 67 (2005) [arXiv:hep-th/0505253].  }
\lref\WittenKon{ E.~Witten, ``On the Kontsevich model and other models
of two-dimensional gravity,'' IASSNS-HEP-91-24}
\lref\IZk{C. Itzykson and J.-B. Zuber, ``Combinatorics of the modular
group II: The Kontsevich Integrals", {\it Int.  Journ.  Mod.  Phys}
{\bf A}, Vol.  7, No.  23 (1992) 5661-5705.  }
\lref\ackm{ J.~Ambjorn, L.~Chekhov, C.~F.~Kristjansen and Y.~Makeenko,
``Matrix model calculations beyond the spherical limit,'' {\it Nucl.\
Phys.}\ {\bf B 404}, 127 (1993) [Erratum-ibid.\ {\bf B 449}, 681
(1995)], \hepth{9302014}.  }
\lref\Bateman{Bateman Manuscript Project: Tables of integral
transforms, vol.  2, chapter XII}
\lref\Martinec{ E.~J.~Martinec, ``The annular report on non-critical
string theory'', hep-th/0305148.}
\lref\bk{M. Bershadski, I. Klebanov, Nucl.  Phys.  B 360 (1991) 559 }
\overfullrule=0pt
%%%
%%%
%%%
%%%
\Title{\vbox{\baselineskip12pt\hbox
{SPhT-T06/086}\hbox{}}}
{\vbox{\centerline
{ Non-Rational   2D Quantum Gravity II. }
\centerline{}
\centerline{  Target Space CFT }
\centerline{ }
\vskip2pt
}}
  \centerline{
I.K. Kostov$^{1}$ and V.B. Petkova$^2$
  }

\vskip 0.7 cm

  \centerline{ \vbox{\baselineskip12pt\hbox
{\it  $^1$ Service de Physique
Th{\'e}orique,
CNRS -- URA 2306,}
  \hbox{ {\it \ \ \ C.E.A. - Saclay,
  F-91191 Gif-Sur-Yvette, France
}
}}}
\medskip
   \centerline{ \vbox{\baselineskip12pt\hbox
{\it  $^2$Institute for Nuclear Research and Nuclear Energy, }
\hbox{\ \  \it 72 Tsarigradsko Chauss\'ee,
1784 Sofia, Bulgaria }}
}

%%%%%%%%%%%%%%%%%%%%%%%%%%%%%%%%%

\vskip 1.5 cm

\baselineskip 11pt
{\ninepoint 
\noindent{
We explore the formulation of non-rational 2D quantum gravity in terms
of a chiral CFT on a Riemann surface associated with the target space.
The CFT in question is constructed as the collective theory for a
matrix chain, which is dual to a statistical height model on dynamical
triangulations.  The heights are associated with the sheets of the
Riemann surface, which represents an infinite branched cover of the
spectral plane.  We consider two examples of height models: the SOS
model and the semi-restricted SOS (SRSOS) model, in which the heights
are restricted from below.  Both models are described in the continuum
limit by theories of 2D quantum gravity with conformal matter,
perturbed by a thermal operator (1,3).  We give a compact operator
expression for the $n$-loop amplitudes as a collection of target space
Feynman rules.  The $n$-point functions of local fields are obtained
by shrinking the loops.  In particular, we show that the 4-point
function of order operators in the SRSOS model coincides with the
4-point function of the ``diagonal'' world sheet CFT studied in
\Bulkcft .
}}

\Date{ }

\baselineskip=11pt
\vfil\eject

\vfill
\eject
%\listtoc\writetoc

\baselineskip=14pt plus 1pt minus 1pt

\newsec{Introduction}

\noindent In a previous paper \Bulkcft\ we studied non-rational
theories of 2D Quantum Gravity (QG), having continuous spectrum of the
matter central charge $-\infty < c_{\rm } <1$.  We explored and
generalized the ground ring approach to the construction of the
tachyon correlation functions.  The method was also tested for an
unconventional model of 2D quantum gravity with interaction which
induces matter charges corresponding to the diagonal of the Kac table.
In this paper we develop a different technique for the computation of
the tachyon correlators based on discrete realizations of the 2D
gravity with non-rational conformal matter.

The simplest statistical model that leads to a non-rational CFT is the
SOS height model, which we will consider on a triangular lattice.  The
local fluctuating variable in the SOS model is an integer height $x$,
which can jump by $\pm 1$ between neighboring sites.  The Boltzmann
weights are complex and depend on two parameters, the ``background
momentum'' $p_0\in [0, 1] $ and the temperature coupling $T>0$.  By
restricting the heights of the SOS model to positive integers we
obtain another height model which we call semi-restricted SOS, or
SRSOS model.  The Boltzmann weights of SRSOS model are real and
anti-symmetric with respect to the reflection $x\to -x$.  As we will
consider the two models alongside, we denote by $\IX$ the target space
of any of them,
 \eqn\defX{ \IX = \cases{ \IZ & for SOS;\cr {\IZ}_+ & for SRSOS.} }
At the rational points the \SRSOS model can be further restricted to
the RSOS models, which have positive Boltzmann weights \refs{\ABF,
\DJKMO}.  The RSOS models form the $A$-series of the $ADE$ Pasquier
models \Pasq.  In this sense the \SRSOS model is ``universal cover''
of the $A$-series of minimal models.  Its target space can be
identified with the Dynkin diagram of the $A_\infty$ Lie algebra,
considered as a certain limit of $A_n$.  It is also possible to
construct the non-rational extension of the $D_n$-series, with
symmetric weights, which we discuss briefly in Appendix C. A common
property of the height models is that they can be mapped to a gas of
self-avoiding and mutually avoiding loops on the dual lattice.  The
observables are introduced by assigning special weights to the
non-contractible loops, while the activity of the contractible loops
  \eqn\defn{ {\rm n} = 2\cos(\pi p_0) }
is determined solely by the background momentum.  The temperature
coupling determines the ``mass'' of the loops.

At the critical temperature $T=T_c$ the large-distance behavior of the
SOS model is that of a gaussian field with curvature term, while the
SRSOS model is described by Virasoro CFT with charge reflection
symmetry.  Both CFT's have the same central charge
\eqn\cdil{ c_{_{\rm critical }}= 1-6\frac{p_0^2}{1+p_0}.  }
There is a set of primary fields which can be defined at microscopic
level.  These are the order operators, which are realized by the
eigenfunctions of the adjacency matrix of the target space $\IX$.
Each order operator is characterized by a momentum $p$.  The
degenerate order operators, placed along the diagonal of the Kac
table, have momenta $p=mp_0$, $m\in\IZ_+$.  The order operators are
well defined also away from the critical temperature, where the long
distance behavior of the height models is described by a perturbation
with the thermal operator $\Phi_{1,3}$ \EY \HM \Gri \FSZ.

The height models coupled to gravity are defined by generalizing the
Boltzmann weights, originally defined on a flat lattice, to an
arbitrary triangulated surface, according to the prescription in
\refs{\Idis, \KPS}.  Then the path integral over two-dimentional
geometries is discretized as a sum over all triangulations with given
topology.

The SOS and SRSOS models have their dual formulations in terms of
doubly infinite or semi infinite matrix chains, similar to the $ADE$
matrix models \adem.  The collective field theory for the matrix chain
is that of a chiral boson on a Riemann surface representing an
infinite branch cover of the spectral plane of the random matrices.
The sheets of the Riemann surface are labeled by the points $x$ of the
target space.  The point on the Riemann surface are enumerated by
pairs $(z, x)$ where $z\in\IC$ is the projection on the spectral plane
and $x\in \IX$ labels the sheet.  Thus each point of the Riemann
surface corresponds to a boundary condition (FZZ-brane), with $z$
being the complex boundary cosmological constant and $x$ being the
height at the boundary.  At the rational points the Riemann surface
represents the algebraic curve of the corresponding minimal model of
2D gravity \SeibergS .

We will be interested only in the genus zero amplitudes, which are
well defined quantities even in a non-unitary theory.  The disk and
the annulus amplitudes are given respectively by the one-point and by
the two-point functions of a free boson on the Riemann surface.  The
amplitudes with more than two boundaries are produced by the
interaction terms.  The latter, associated with the branch points, are
needed to preserve the conformal invariance \Iconfm.  The operator
solution of the Virasoro constraints yields a set of Feynman rules
for calculating the loop correlation functions, which allows to
evaluate any such function by a finite sum over all possible
intermediate channels.  In the case of rational $ADE$ string theories
these rules were formulated in \Higkos.

We work out explicitly the example of the 4-point function and will
compare the result to the prediction of the world-sheet CFT. In the
case of the SOS model, our result reproduces at the critical point the
Di Francesco-Kutasov formula \DiK\ obtained for gaussian matter field
with curvature term.  In the case of the SRSOS model, our result for
the 4-point function matches the one obtained in \Bulkcft\ for the
``diagonal'' perturbation of Liouville gravity.  This perturbation is
generated by the Liouville-dressed vertex operator of dimension zero
obtained from the identity by charge reflection.

The paper is organized as follows.  In sect 2 we give the definition
of SOS and SRSOS models on a triangulated surface with curvature
defects and their representation in terms of a gas of loops.  The loop
gas representation of the correlation functions of order operators is
sketched in Appendix A. In sect.  3 we reformulate the height models
on dynamical triangulations in terms of doubly or semi infinite matrix
chains.  In sect.  4 we give the solution for the disk amplitude in
the continuum limit.  In sect.  5 we construct the target space CFT as
a bosonic field theory on the Riemann surface associated with the disk
amplitude.  The Feynman rules for calculating the $n$-loop amplitudes
are obtained in sect.  6.  In sect.  7 we work out the example of the
4-point function of order operators.  Summary of the results is given
in sect.  8.

\newsec{The SOS and SRSOS models on a triangulation with curvature
defects}

\subsec{Local Boltzmann weights and mapping to a loop gas}

\noindent Let $\CG$ be a triangulated surface with the topology of a
sphere with $n$ boundaries.  The triangulation is characterized by its
vertices $r$, its links $l=<r_1r_2>$ and its triangles
$\Delta=<r_1r_2r_3>$ which are assumed equilateral.  The curvature
defects are associated with the points $r\in \CG $ with coordination
number $c_r$ (the number of triangles having this point as a vertex)
different than 6.  The local curvature at such point is equal to the
deficit angle,
\eqn\currb{ \hat R_r= \frac{2\pi}{3} (6-c_r).  }
 Similarly, the boundary curvature at the points $r\in\p\CG$ is
\eqn\bcurrb{ \hat K_r= {\pi\over 3} (3-c_r).} By Euler's relation the
total curvature is \eqn\diseul{ \sum_{r\in\CG} \hat R_r + 2\sum_{r\in
\p\CG} \hat K_r = 4\pi (2 - n).  }

The SOS and SRSOS models on the triangulation $\CG$ are defined as
follows.  To each node $r\in \CG $ we associate an integer height $x
_r\in \IX$, where the target space $\IX$ is defined in \defX. The
allowed height configurations are such that the heights of two
nearest-neighbor points are either equal or differ by $\pm 1$.  If one
thinks of the target space as a one-dimensional graph (Fig.  1), then
the allowed height configurations are the maps $\CG\to \IX$ which
preserve the nearest neighborliness, {i.e.}, such that points are
mapped to points and links are mapped either to points or to links.
Since the space $\IX$ is discrete, the momentum space $\IP$ is
compact: $p\sim p+2$.  In the case of SRSOS there is additional
reflection symmetry, $p\sim-p$.

%%%%%%%%%%%%%%%%%%%%%%%%%%%%%%%%%
  \epsfxsize=180pt
\vskip 20pt
{\centerline{ \epsfbox{ 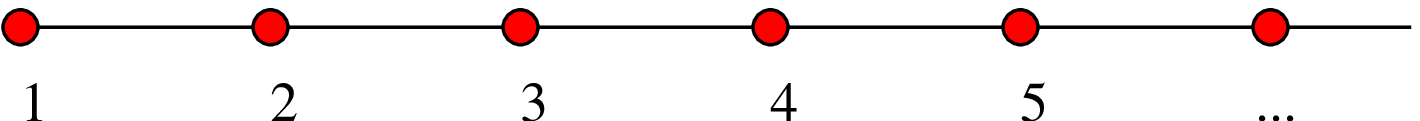 }}}
\vskip 5pt
%%%%%%%%%%%%%%%%%%%%%%%%%%%%%%%%%%

\centerline{\ninepoint Fig.  1: The target space of the SRSOS model
as a one-dimensional graph.  }

  \vskip 10pt

\noindent The partition function $Z _{\CG}(x_1,..., x_n)$ is the sum
over all allowed maps $ \CG\to \IX$ such that the boundaries have
fixed heights $x_1,..., x_n$.  Besides the standard local factors
$W_\Delta$ associated with the faces of $\CG$, there are extra weights
$W_\bullet$ that come from the curvature defects:
\eqn\pfcmt{Z _{\CG}(x_1,..., x_n) = \sum_{ \CG\to \IX}\
\prod_{<r_1r_2r_3> } W_{\Delta}(x_{r_1},x_{r_2},x_{r_3})\ \prod _{ r}
W_{_{\bullet}}(x_r).  }
The Boltzmann weights are expressed in terms of the function
	\eqn\SSxx{\Sx = \cases{ \exp(-i\pi p_0 x)& for SOS;\cr \sin
	(\pi p_0 x) & for \SRSOS } }
which plays the same role as Peron-Frobenius vector in the $ADE$
height models.  It is an eigenvector of the adjacency matrix of the
graph $\IX$ with eigenvalue $2\cos(\pi p_0)$.  The background momentum
$p_0\in [0,1]$ is assumed in this paper to be non-rational.  By
definition, the weight of a triangle $\Delta_{123}$ with heights $x_1,
x_2, x_3$ is invariant under cyclic permutations and is non-zero only
if the heights of each pair of points are either the same or adjacent.
The possibilities are either $x_1=x_2=x_3$ or $x_1=x_2 = x_3 \pm 1$,
up to a cyclic permutation, and have weights
\eqn\Wtr{\eqalign{ W_{\Delta} (x,x,x )= 1;\qquad W_{\Delta} (x,x,x\pm
1)= {1\over T} \( {S_{x\pm 1} \over S_{x} }\) ^{1/6} , }}
where $T$ is a positive constant called temperature.  The weight
associated with the vertex $r\in \CG$ with curvature defects are
\eqn\Wrp{ W_{_{\bullet}}(x_r)=\cases{ (S_{x_r})^{ \hat R_r/4\pi } & if
$r\in \CG$;\cr
 % &\cr
    (S_{x_r})^{ \hat K_r/2\pi } & if $r\in \p\CG$} .  }
While Boltzmann weights of the SOS model are complex for any $p_0\ne
0$, those of the \SRSOS model are real, but not always positive.

The height model is also described in terms of a {\it loop gas}.  Let
us represent the weights of the triangles with admissible heights
symbolically as
% %
\epsfxsize= 140pt \eqn\tttrfig{\epsfbox{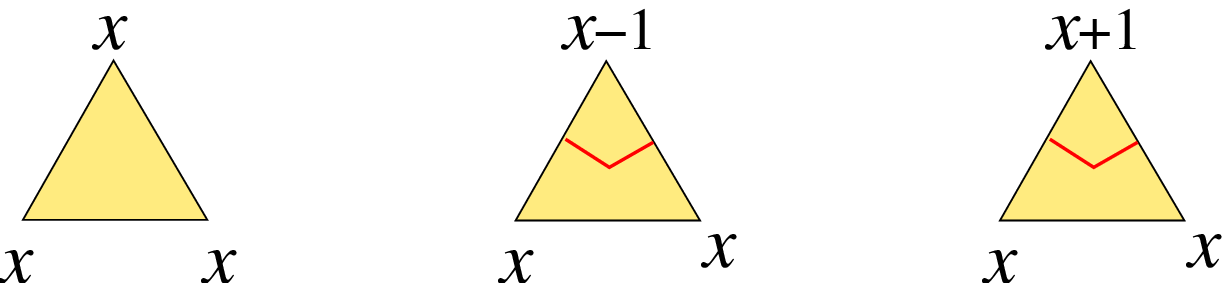}}
That is, whenever the three heights round the triangle are not the
same, a line separates the two vertices with the same height from the
vertex with different height.  The line is made by two segments at
angle $\pi/3$, orthogonal to the edges they start from.  Then each
admissible height configuration determines a pattern of closed
nonintersecting loops (Fig.  2).  With our choice of boundary
conditions the domain walls do not cross the boundary.

%

%%%%%%%%%%%%%%%%%%%%%%%%%%%%%%%%%
  \epsfxsize=140pt
\vskip 20pt
\hskip 85pt
\epsfbox{ 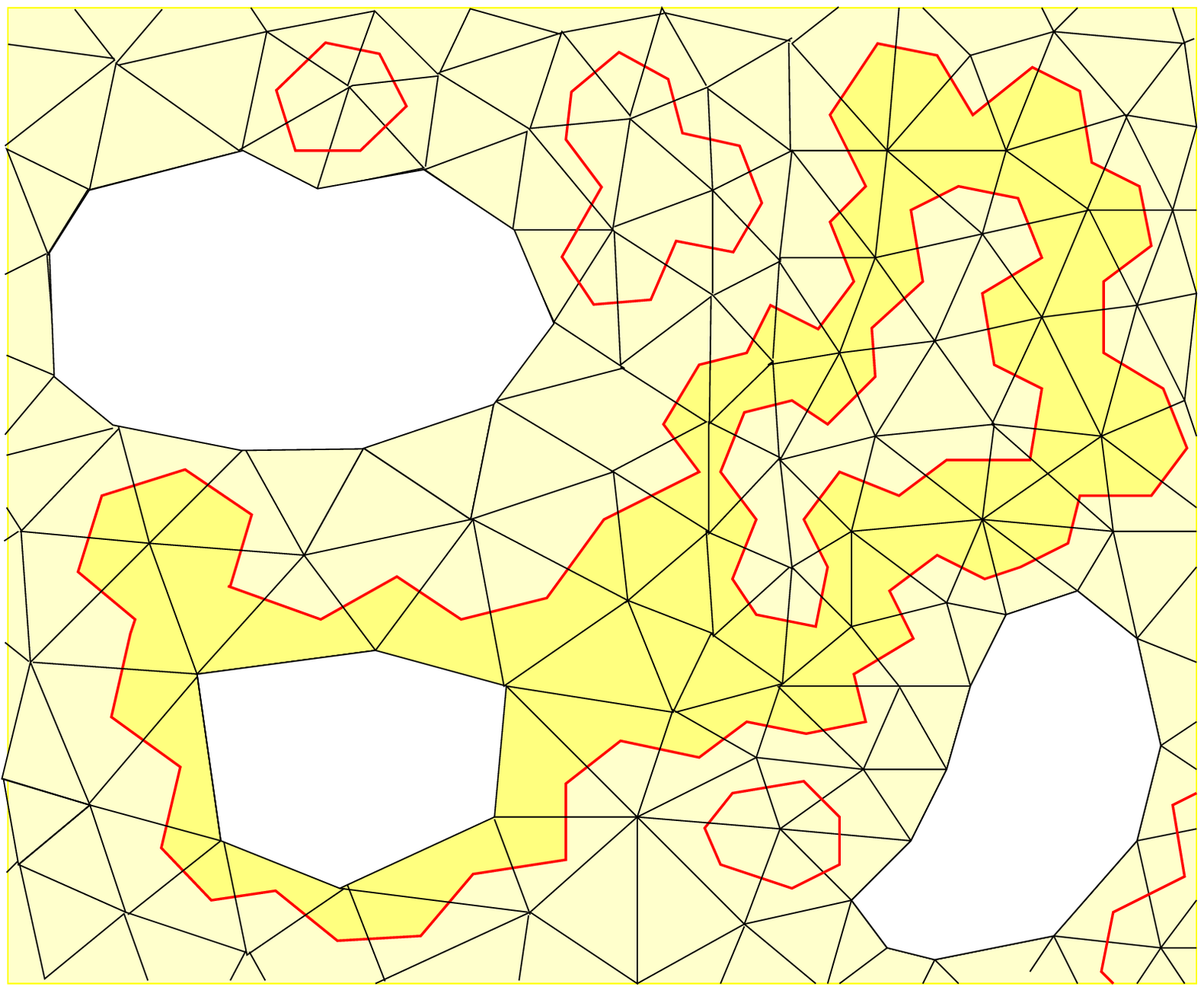   }
\vskip 15pt
%%%%%%%%%%%%%%%%%%%%%%%%%%%%%%%%%%

 \centerline{\ninepoint \vbox{ \hbox{Fig.  2: A loop configuration on
 a triangulation with boundaries.  The boundary of} \hbox{ the
 highlighted domain has $n_{\rm b}=3$ components, two of them being
 dynamical loops.  } }}

  \vskip  10pt

\noindent The Boltzmann weights in \pfcmt\ are such that the weight of
each loop gas configuration factorizes to a product of the weights of
the connected domains of constant height.  The weight of a domain
$\CD$ is topological -- it depends on the height $x$ of the domain
only through its Euler characteristics, {\it i.e.,$\!$} the number of
its boundaries:\foot {In the case of a lattice with higher genus,
there would be also a factor $(S_x)^{-2}$ for each handle trapped in
the domain.}
\eqn\weightD{ \O_{\CD}(x)= (\Sx )^{2-n_b}, \quad n_b = \#\ {\rm
boundaries \ of} \ \CD .  }

\noindent The proof goes as follows.  We can distribute the weights of
the triangles that belong to two different domains, given by the
second term of eqn.  \Wtr, in such a way that the weight of each domain
$\CD$ is of the form $(S_x)^q$, where $x$ is the height of the domain.
It remains to evaluate the power $q$.  It receives contributions from
the vertices in the bulk inside the domain wall, from the vertices of
the boundaries of $\CG$ (if any) that are also boundaries of $\CD$,
and from the triangles along the domain walls that delimit $\CD$.  If
there are no domain walls at all, that is, $\CD=\CG$, then the total
power of $S_x$ is $q=2- n$ according to the Euler relation \diseul.
If there are one or more domain walls, each domain wall contributes a
factor $S_x^{\hat K}$, where $\hat K$ is the total boundary curvature
along it.  Indeed, the weight of a domain wall is a product of factors
$(S_x)^{\pm 1/6}$ that come from the triangles along it and account
for the defects $\pm \pi/6$ along its edge.  Applying again the Euler
relation we find $q=2-n_b$, where $n_b$ is the total number of
(connected) boundaries of the domain, given by the number of
boundaries of $\CG$ trapped in the domain plus the number of the
domain walls.

     \smallskip

As a consequence, the partition function \pfcmt\ can be reformulated
as a sum over all configurations of self and mutually avoiding loops
on the dual lattice and heights $ x_{\CD}$ associated with the
connected domains $\CD$:
\eqn\lpgz{ Z _{\CG}( x_1,...,x_n)= \sum_{\ \ \ \ ^{\rm loop \
configu-}_{\rm rations\ on\ \CG}} \ T^{-\Ltot} \ \sum_{{\rm heights}\
x_{\CD}} \ \prod_{\CD} \O_{\CD}(x_{\CD}) .  }
Here $\Ltot$ is the total length of the loops, which is equal to the
total number of loop segments.  The sum over the heights of the
domains, $\{x_{\CD}\}$, is such that the heights of the neighboring
domains differ by $\pm 1$.  In this way  the mapping to a loop gas, originally
established for a flat lattice \nienhuis, remains true also in the
case of a lattice with curvature defects.

For the disk partition function $(n=1)$ the one can perform readily
the sum over the heights but one using repeatedly the relation
\eqn\SAx{ \sum_{x'}
A_{xx'}S_{x'} = 2\cos\pi p_0\, S_x.  } 
In this way the sum over heights produces a factor $ (2\cos\pi
p_0)^{\rm \#\ loops}$ times a factor $S_{x_{1}}$ depending on the
boundary height $x_{1}$.  Up to this last factor, this is the disk
partition function of the $O({\rm n})$ loop gas \nienhuis\ with
activity ${\rm n}=2\cos\pi p_0$ per loop.

In the case of more than one boundary, the sum over the heights can be
performed by Fourier transformation with the eigenvectors of the
adjacency matrix \eqn\defSpx{ S^{(p)}_x= \cases{e^{i\pi p x} & for
SOS;\cr \sin (\pi p x) & for SRSOS. } } In this case the loops can
have different activities.  The contractible loops, {\it i.e.}, the
loops for which there is no topological obstacle for shrinking them to
points, have activity $2\cos\pi p_0$.  A non-contractible loop
contributes a factor $ (2\cos\pi p)$, where $p\in\IP $ is the momentum
associated with the loop.  In Appendix A we give the loop gas
representation of the correlation function of $n$ order operators on
the sphere, which is the Fourier transform of the the partition
function \lpgz\ with boundaries of length zero.

\subsec{Phase diagram and critical points }

\noindent The critical thermodynamics of the SOS and \SRSOS models is
that of the $O({\rm n})$ loop gas with ${\rm n}= 2\cos(\pi p_0)$.  The
phase diagram (Fig.  3) of the $O({\rm n})$ model on the the honeycomb
lattice (dual to the regular triangulation) was first established in
\nienhuis .  At the critical temperature $T_c = 2\cos
\frac{\pi}{4}p_0$ the loop gas model is solvable and is described by a
CFT with central charge
\eqn\cdil{ c_{_{\rm critical }}= 1-6{p_0^2\over 1+p_0}.  } 
The critical point is also known as the {\it dilute phase} of the loop
gas.  For $T>T_c$ the theory has a mass gap.  The low-temperature, or
{\it dense}, phase $T<T_c$ is a flow to an attractive fixed point
\BNh\ at $T^{\rm dense}_c = 2\sin \frac{\pi}{4} p_0,$ where the theory
is again solvable and is described by a CFT with smaller central
charge
\eqn\cden{ c_{_{\rm dense} }=1-6 {p_0^2\over 1-p_0}.  }

%%%%%%%%%%%%%%%%%%%%%%%%%%%%%%%%%
\epsfxsize=130pt
\vskip 20pt
\hskip 100pt
\epsfbox{ 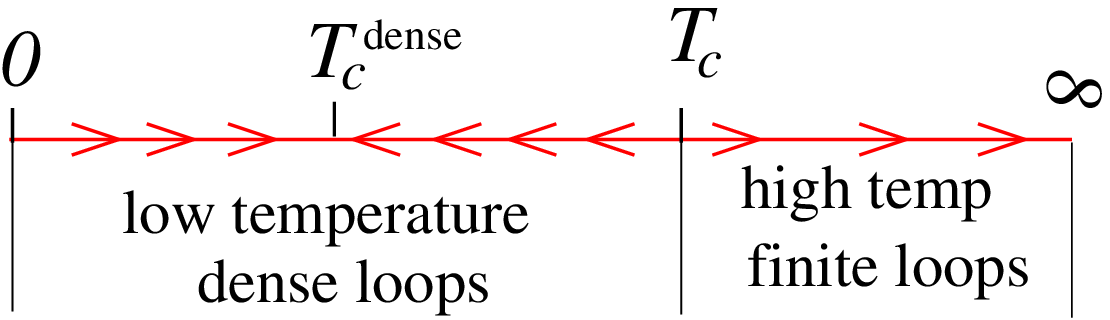   }
\vskip 5pt
%%%%%%%%%%%%%%%%%%%%%%%%%%%%%%%%%%
\centerline{\ninepoint \vbox{\hbox{Fig.  3: Phase diagram of the loop
gas on a regular (honeycomb) lattice.} } } \vskip 10pt

\noindent The scaling behavior of the model in the vicinity of a
critical point is described by an action of the form \FSZ
\eqn\aaca{\CA =\CA_{\rm critical} + \delta T \int \Phi_{1,3} }
where $\delta T = T-T_c $ and $ \Phi_{1,3}$ is the thermal operator
with conformal dimensions \eqn\dtemp{ \Delta_{1,3}= \bar \Delta_{1,3}
= {1-p_0\over 1+p_0}.  }
When the thermal operator $\Phi_{1,3}$ is added to the action, it
generates a tension of the loops proportional to $\d T$.  For $ \delta
T >0$ the deformation \aaca\ describes, going from short to the long
distance scales, the flow to a massive theory with mass gap
 \eqn\mssc{ m \sim \delta T ^{1/(2-2\Delta_{1,3})}= \delta T
 ^{1+p_0\over 4p_0} .}
When $\delta T <0$ the deformation \aaca\ describes a massless flow
between two different CFT with central charges \cdil\ and \cden.  In
the CFT for the dense phase, describing the IR limit, the flow to the
attractive critical point is generated by another, irrelevant,
operator, $\Phi_{3,1}$.

 \subsec{Mapping to a  gaussian field}

 \noindent $\bullet$ SOS model

It has been argued that at the critical points the large distance
behavior of the SOS model on a lattice with curvature defects is that
of a gaussian field $\chi(\s_1,\s_2)$ with a background charge
proportional to the local curvature \refs{\nienhuis, \FodaIN}.  The
product of the factors \Wrp\ associated with the curvature defects
yields the curvature-dependent term in the effective action
\eqn\actfr{
 \CA _{\rm gauss} = \frac{1}{4\pi} 
\int d^2\s \[ (\nabla \chi )^2 + i e_0\chi\].
}
The two critical points are described by different values of the
background charge $e_0$, related to the background momentum $p_0$ by
\eqn\chihash{ \eqalign{ e_0 = \pm {p_0 \over \sqrt{1\pm p_0}},\ \ 
{\rm with} \ \  \qquad ( {\ninepoint + {\rm \ for\ dilute},
  \ - \ {\rm for\ dense} } )
  .}}
In the SOS model the order operators $\psi_p(x) = S^{(p)}_x/S_x$ have
the form of plane waves
  \eqn\orderp{\psi_{p}(x) = e^{ i \pi (p_0-p)x} = e^{2\pi i q x}.}
At each of the critical points they can be identified with vertex
operators of the gaussian field with conformal dimensions
 \eqn\DimvSOS{ \Delta_{p}= {p^2- p_0^2\over 4(1\pm p_0)} =
 {q(q-p_0)\over 1\pm p_0} \qquad ( {\ninepoint + {\rm \ for\ dilute},
 \ - \ {\rm for\ dense} } ).}
  
\noindent $\bullet$ \SRSOS model

In this case the correlation functions are constructed from vertex
operators and screening charges.  To get an intuition about how the
vertex operators appear in the microscopic theory, let us notice that
in the \SRSOS model the effect of the curvature does not reduce to a
pure phase factor, as it was the case for the SOS model.  Here we
follow the argument of \FodaIN. After formally expanding, with $\Sx $
given by \SSxx, one gets
  $$
   \log \Sx =-i \pi p_0 x - e^{2\pi i p_0 x} + ..  . 
  $$
The exponential term corresponds to a vertex operator \orderp\ with
$q= p_0 + $ integer.  The choice $q=p_0+1$ gives one of the screening
charges added to the action \actfr, $e = \sqrt{1+p_0}$.

\newsec{ SOS and  \SRSOS models on dynamical triangulations}

\noindent When considered on dynamical triangulations, the heights
models give the desired solvable discrete models of non-rational 2D
quantum gravity.  The partition function in the ensemble of
triangulations with the topology of a sphere with $n$ boundaries, or
shortly $n$-loop amplitude, is defined as
\eqn\Zgrav{Z_{(L_1, x_1), (L_2, x_2), ..., (L_n, x_n)}
=\sum_{\CG_{L_1, \dots, L_n}} \k ^{A} \ Z_{\CG_{L_1, \dots, L_n}}(x_1,
\dots, x_n).  } 
The sum goes over all triangulations $\CG_{L_1, \dots, L_n}$ with the
above topology and fixed lengths (number of edges) $L_1, \dots , L_n$
of the boundaries, and the ``cosmological constant'' $\k$ coupled to
the area $A$ (the number of triangles) of the triangulation.

\subsec{Dual description in terms of doubly infinite or semi infinite
  matrix chains}

\noindent There is a dual, matrix, formulation of the height models on
dynamical triangulations in terms of a doubly infinite (for SOS) or
semi-infinite (for SRSOS) matrix chains.  These matrix models are
similar to the $ADE$ matrix models \adem, with some subtleties related
to the fact that they describe non-unitary theories.  The fluctuating
variables are hermitian $N_x\times N_x$ matrices ${\bf M}_x$ and the
complex rectangular $N_x\times N_{x+1}$ matrices ${\bf C}_{x }$ \
($x\in \IX$).  The hermitian matrices ${\bf M}_x$ are naturally
associated with the sites $x$ while the complex matrices ${\bf C}_{x }
$ are associated with the links $<x, x+1>$ of the target space.
The partition function of the matrix chain is given by the integral
   \eqn\mpatr{\CZ[V]= \int [d{\bf M]}\, [d {\bf C}][ d {\bf
   C}^{\dagger}]\ \exp\( - {1\over\gs} \CA\),}
\eqn\Mact{\eqalign{ \CA& = \sum_{x\in \IX} {S_x }\, \tr \Big( \hf {\bf
M}_{x} ^2- \frac{1}{3} \k \, {\bf M}_{x} ^3 +T {\bf C}^{\dagger}_{x
}{\bf C}_{x } - \k{\bf C}_{x }{\bf C}^{\dagger}_{x }{\bf M}_{x }
-\k{\bf M}_{x+1}{\bf C}^{\dagger}_{x} {\bf C}_{x }} \Big).  }
The perturbative expansion of the matrix integral is constructed
according to the Feynman rules in Fig.  4, which correspond to the
five terms in the action \Mact.  The correlation functions in the
matrix model can be expressed in terms of fat (double lined) graphs.
We will be interested in the limit $N_x\to \infty$ for all $x\in \IX$,
in which limit only the planar graphs survive.  Each planar graph is
in one-to-one correspondence with a triangulated surface.  The
vertices of the triangulation (dual to the faces of the planar graph)
have label $x\in\IX$.  Thus the sum over planar diagrams reproduces
the path integral for the height model in the ensemble of
triangulations.

%%%%%%%%%%%%%%%%%%%%%%%%%%%%%%%%%
\epsfxsize=200pt
\vskip 20pt
\centerline{\epsfbox{ 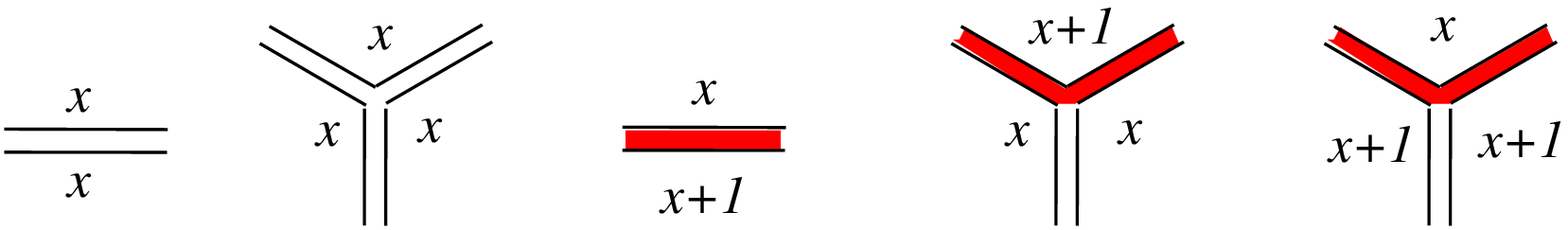   }}
\vskip 5pt
%%%%%%%%%%%%%%%%%%%%%%%%%%%%%%%%%%

\centerline{\ninepoint Fig.  4: Feynman rules for the \SRSOS matrix
model }

\vskip  10pt

\noindent In order to obtain the correct Boltzmann weights, we must
tend the size of the matrices to infinity in a particular way:
\eqn\Nxx{ {\gs N_x } \to S_x , \qquad \gs\to 0.  } Using the
interpretation of the Feynman graphs as triangulated surfaces, one can
easily see that the $n$-loop amplitude \Zgrav\ is equal to the genus
zero connected correlation function of a product of $n$ traces
\eqn\partfZn{ Z_{(L_1, x_1), (L_2, x_2), ..., (L_n, x_n)} =\lim
_{\gs\to 0}\ \gs^{2-n} \<\!\!\!\!\< \prod _{i=1}^n \tr ({\bf M
}_{x_i}^{L_i})\> \!\!\!\!\> \, .  }

The reader might object that the limit \Nxx\ does not exist, because
$S_x$ is not a positive number.  The limit is in fact well defined
only at the rational points of the SRSOS model, $p_0= 1/(h+1)$, with
$h\in \IZ_+$.  In this case $S_x$ is Peron-Frobenius vector for the
adjacency matrix of the $A_{h}$ Dynkin graph and all its components
are positive.  However, we can impose the condition \Nxx\ in a weaker
sense as an analytical continuation from positive integer values.

As the observables \partfZn\ are associated with the ${\bf
M}$-matrices, we can integrate out the ${\bf C}$-matrices in the
partition function.  After the integration the partition function
\mpatr.  can be written, up to a factor equal to the volume of the
symmetry group $\otimes _{x\in \IX} SU(N_x)$, as an integral with
respect to the eigenvalues $\l_{i,x}, \ i=1, ..., N_x,$ of the
hermitian matrices ${\bf M}_x$:
   \eqn\prtFAr{ \CZ \sim \int \limits_{-\infty}^{\infty} \prod _{x\in
   \IX} \prod_{ i=1}^{ N_x } d\l_{i,x} \, e^{-{1\over\gs} S_x ({1\over
   2} \l_{i,x}^2 - {1\over 3} \k \l_{i,x}^3)} \ \prod_{i<j}^{N_x}
   (\l_{i}^x-\l_{j}^x)^2 \prod_{ i, j} {1\over |T - \k(
   \l_{i,x}+\l_{j,x+1})|}.  }
The explicit dependence on the second coupling $\k$ can be eliminated
by a linear change of variables
\eqn\shiftM{ {\bf M}_x \to \frac{1}{ \k } {\bf M}_x + \frac{T}{2 },
\qquad \gs \to \k^2 \gs.
} 
After that the partition function \prtFAr\ takes the form
\eqn\prtFAr{\eqalign{ \CZ &\sim \int \limits_{-\infty}^\infty \prod
_{x, i} {d\l_{i,x} }\ e^{ - {1\over \gs} V_x(\l_{i,x})} \ {\prod _{x;
i< j} (\l_{ i,x} - \l_{j, x})^2\over \prod_{i,j} |\l_{i, x}+ \l_{j,
x+1}|} } }
where the cubic potential depends only on the temperature $T$,
\eqn\potv{V_x(z) \equiv S_x V(z), \qquad V(z) = \frac{T(2 - T)}{4} z+
\frac{1- T}{2} z^2 - \frac{1 }{3} z^3 .  }
The coupling $\k$, the lattice cosmological constant, reappears in the
definition of the planar limit, originally given by \Nxx, which now
becomes
\eqn\Nxxbis{ {\gs N_x } \to \k^2 S_x , \qquad \gs\to 0.  }

The eigenvalue integral \prtFAr\ is not convergent.  To make it
convergent, we will drop the absolute value in the denominator and
will understand the integrals over the eigenvalues as contour
integrals in the complex plane.  The new integral will have the same
perturbative expansion as \prtFAr.  We choose the integration contour
$\CC$ so that it starts at $\l_{i,x}= -\infty$, passes near the origin
and then goes to infinity along a direction in which the potential
grows (Fig. 5a).  This is a standard prescription used to define
rigorously the matrix integrals that appear in 2D quantum gravity, see
e.g. \David .  To avoid the poles of the integrand at $\l_{x,i} = -
\l_{x',j}$ we also require that the contour $\CC$ does not intersect
the contour $\bar \CC$ obtained from $\CC$ by a reflection $z\to - z$.
 
 %

%%%%%%%%%%%%%%%%%%%%%%%%%%%%%%%%%
\vskip 20pt 
\centerline{\epsfxsize=150pt  \epsfbox{  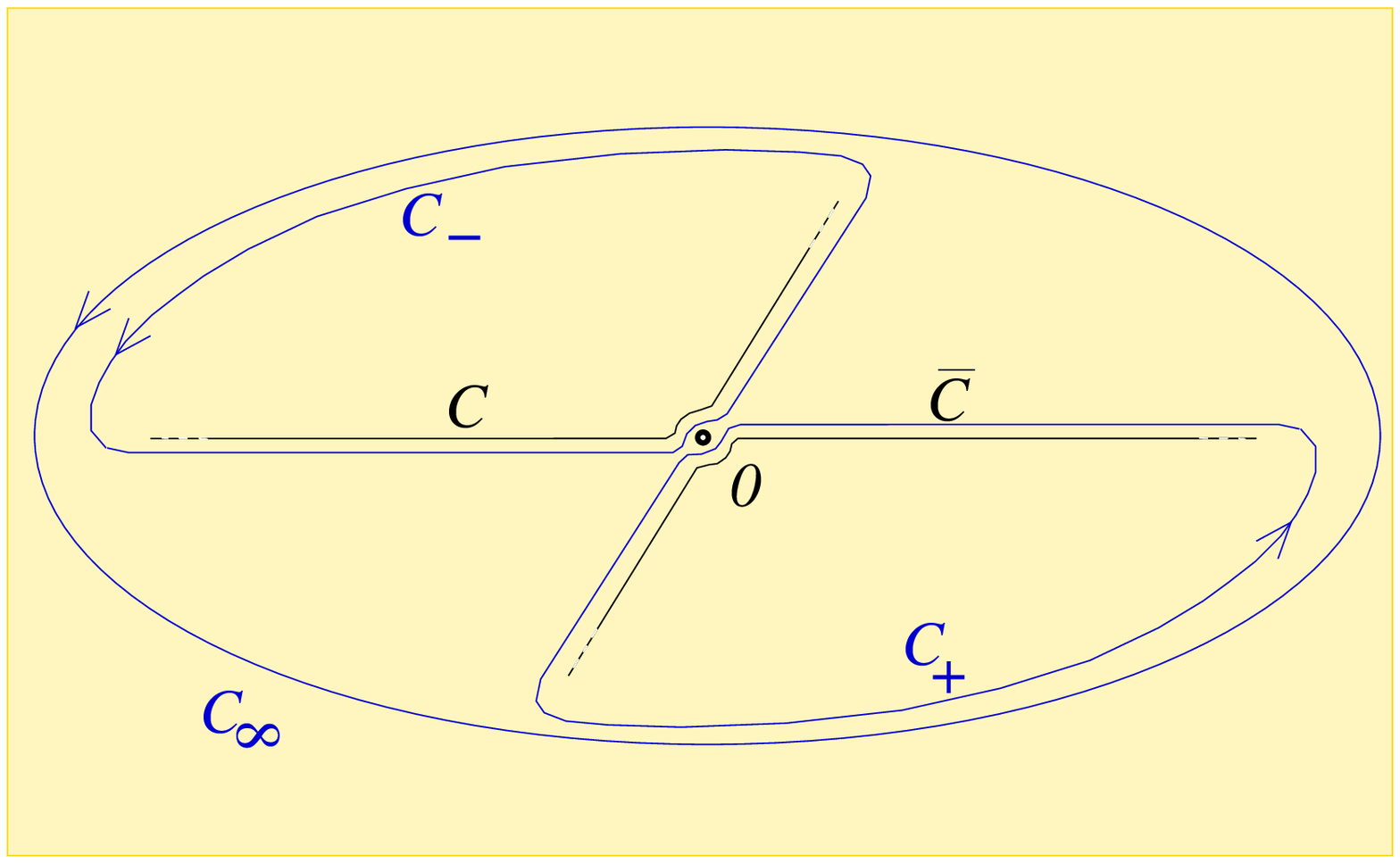} 
\hskip 0.7cm \epsfxsize=150pt
\epsfbox{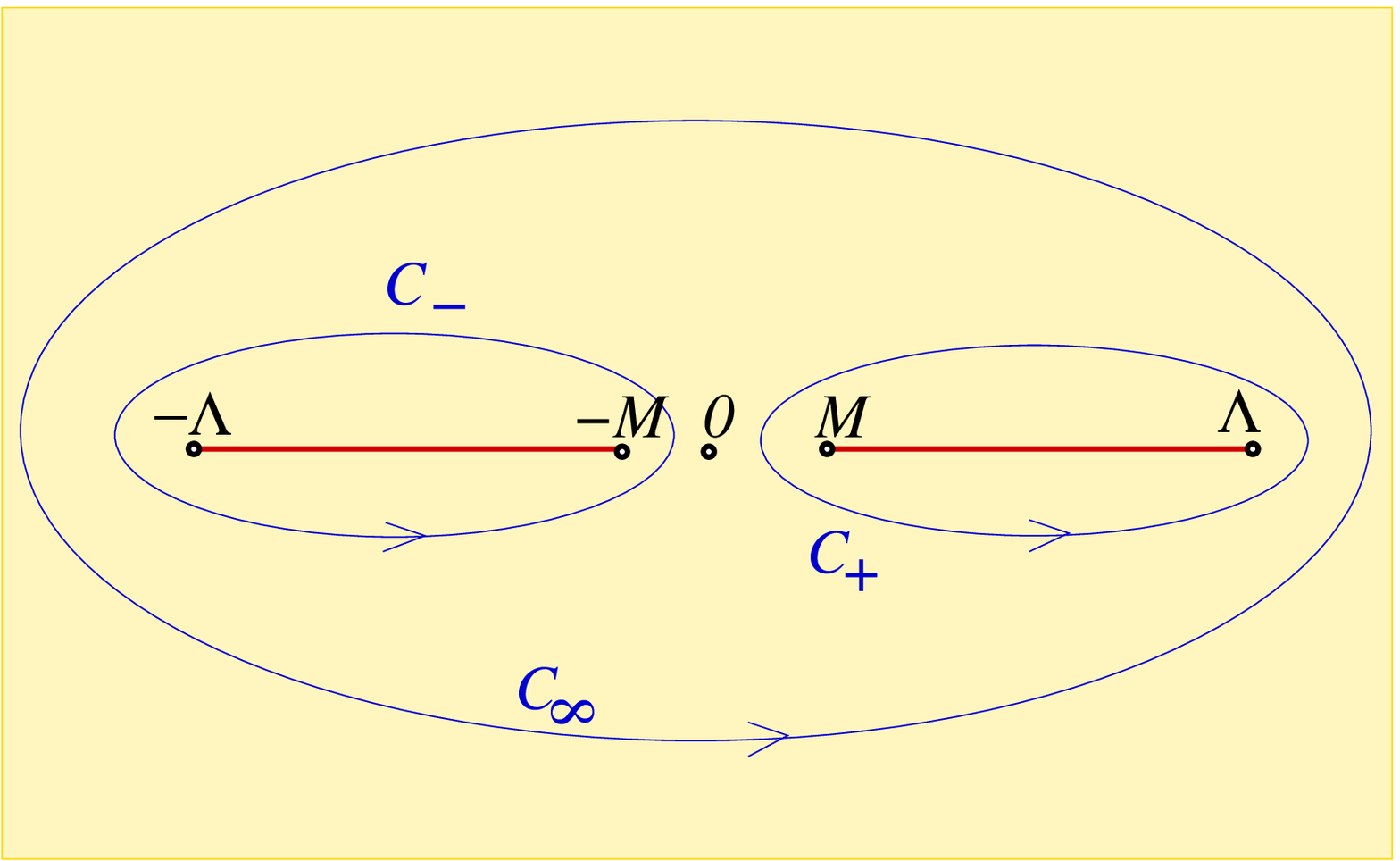   } }

\centerline{a\hskip 6cm b}

%%%%%%%%%%%%%%%%%%%%%%%%%%%%%%%%%%

\vskip 5pt

\centerline{\ninepoint \vbox{\hbox{Fig.  5a.  Integration contours in
the complex plane of the eigenvalues for cubic potential} \hbox{Fig.
5b.  Integration contours relevant for the quasiclassical limit} } }

\vskip  10pt

\noindent There are several choices for the second branch of the
contour $\CC$, which however lead to the same perturbative expansion.

\subsec{The matrix chain  as a chiral conformal field theory}

\noindent It is known that a large class of matrix models can be
reformulated as conformal field theories \MMM \CMM \Morozov \Icar
\Iconfm \IHouches .  The correspondence MM$\leftrightarrow$CFT can be
established if the measure in the eigenvalue integral can be produced
by screening charge operators.  We will show that this is the case for
the integral \prtFAr\ where the integration goes in the complex plane
along the contour $C$.

We associate with each point $x\in\IX$ a chiral boson $\vp _x(z)$ 
two-point correlator  
\eqn\ccrZ{\< 0| \vp_x(z)\vp_{x'}(z')|0\>=\delta_{x,x'}\, \log(z-z'),
} 
where the right/left vacuum is annihilated by the negative/positive
power part of the Laurent expansion of the fields $\p\vp_x(z)$ in $z$.
We introduce two more bosonic fields,
\eqn\defGa{\eqalign{
\G_x(z) &=   \vp_{x }(z) - \vp_{x+1}(-z),
}}
and the conjugated field  $\Phi_x(z)$, defined by
   \eqn\PhPh{ \< 0| \p_z \Phi_x(z) \G_{x'}(z') |0\> ={ \delta_{x,x'}
   \over z-z'}.  }
The field $\Phi_x(z)$ is related to $\vp_x(z)$ and $\G_x(z)$ by
\eqn\defPh{
\Phi_x(z) - \Phi_{x+1}(-z) = -\vp_{x+1}(-z) , %\ \ x\in\IX, 
}
\eqn\dualb{
  2\Phi_x(z) -  \sum_{x'} A_{xx'} \Phi_{x'}(-z)= \Ga_x(z)  ,
}
where $A_{xx'}$ is the adjacency matrix of $\IX$.

It is straightforward to check that the integrand of \prtFAr\ can be
written as Fock space expectation value of a product of vertex
operators
\eqn\defE{ E_x(z) =  :e^{-\vp_x(z) } : \ : e^{\vp_{x+1}(x)  } :
.  } 
Let $\langle \vec N | $ be the charged vacuum state defined by the
asymptotics \eqn\asinf{ \p \Phi_x(z) = - {N_x \over z } + \dots ,
\qquad z\to\infty.  }
Then  using the OPE of the vertex operators one finds
 \eqn\intgrd{ \prod_{x\in \IX} {\prod _{i\ne j} (\l_{ i,x} - \l_{j,
 x})\over \prod_{i,j} (\l_{i, x}+ \l_{j, x+1})} = \langle \vec N|
 \prod _{x, i} \ E_x(\l_{i,x})|0 \rangle.  }
To generate the external potential it suffices to replace the left
vacuum by a coherent state
  \eqn\leftV{ \langle B|=\langle \vec N | \exp( \sum_{x\in\IX}
  H_x),\qquad H_x={1\over \gs} \oint\limits_{C_\infty} {dz\over 2\pi
  i} V_x(z) \p \Phi_x(z) }
where the contour $\CC_\infty$ encircles both $\CC$ and $\bar\CC$
(Fig.  4a).  This is easily verified by applying the operator product
expansion \PhPh\ 
%  %
%  $$\p \Phi_x_x (z) E_{x'}(z') = \d_{x,x'} (z-z')^{-1} E_x(x) + ...
%  $$ 
%  %
to commute $e^{H_{x}}$ to the right.  The result is a factor
$e^{-{1\over\gs}V_x}$ per eigenvalue.

To write the Fock space representation of the eigenvalue integral
\prtFAr\ we define the screening operators
\eqn\defscrop{ Q_x = \int_{\CC} {dz }
\, E_x(z) .  } 
Then we can write the eigenvalue integral in the form of a scalar
product \eqn\FockR{ \CZ = \< B|\Omega\>, } where $ \langle B|$ is
defined by \leftV\ and
\eqn\defOm{ \big| \Omega\big\rangle = \exp\big( \sum_{x\in \IX} Q_x
\big) \big| 0\big\rangle .  }

   \bigskip

\rb {\it Correspondence between the observables in CFT and the matrix
model }

The expectation value $\langle \ \ \rangle_{_{\rm MM}}$ in the matrix
model becomes the vacuum expectation value in terms of Fock space.  To
any observable $\CO_{_{\rm MM}}$ in the matrix model one can associate
an operator $\CO $ such that
\eqn\expF{ \< \CO_{_{\rm MM}}\> _{_{\rm MM}}= {\big\langle B \big |
\CO \big| \Omega\big\rangle \over \big\langle B \big |
\Omega\big\rangle} \equiv\< \CO\> .  }
In particular, the resolvent of the matrix ${\bf M}_x$ is represented
in the CFT as
 \eqn\WPhi{\eqalign{ W_x(z)\equiv {g_s}\, \tr\({1\over z- {\bf M}_x}\)
 \ \leftrightarrow \ - \gs \, \p\Phi_x^{[+]}(z) \,, }}
where $\p\Phi^{[+]}(z)$ is the negative part of the Laurent series of
$\p\Phi(z)$, which annihilates the right vacuum, $\p\Phi^{[+]}|0\ra
=0$.

In other words, the resolvent is equal to (minus) the singular at
$z=0$ part of Laurent series of $\p\Phi(z)$, which annihilates the
right vacuum, $\p\Phi^{[+]}|0\ra =0$.

The other important observable is the FZZ-brane, which measures the
effect of suppressing the integral with respect to one of the
eigenvalues of the matrix ${\bf M}_x$ at assigning to it a complex
value $z$.  This matrix observable corresponds to the vertex operator
\defE,
\eqn\phreZ{ e^{-{1\over\gs}V_x(z)} \det (z-{\bf M}_x)^2\ \prod_{x'}
\det(z+{\bf M}_{x'}) ^{-A_{xx'}} \ \leftrightarrow\ E_x(z)= :
e^{-\G_x(z)}:\ .}
In the limit $\gs\to\infty$ we can replace $\< : e^{-\G_x(z)}:\> =
e^{-\<\G_x(z)\>}$.  Thus the expectation value of the field \dualb\
gives the effective potential of one eigenvalue taking value $z$ in
the mean field produced by the other eigenvalues,
\eqn\dualba{ {1\over\gs}V_x(z)- 2\tr \log (z-{\bf M}_x)
+\sum_{x'}A_{xx'} \tr\log (z+{\bf M}_{x'})\ \leftrightarrow\ \G_x(z).
}

\noindent
\rb {\it Virasoro constraints}

In the matrix model formulation, the $n$-loop amplitudes satisfy a set
of loop equations that follow from the translational invariance of the
matrix integration measure.  The loop equations have been first
derived from the combinatorics of planar graphs \Idis.  They have the
form
\eqn\ADEeqa{\Big\langle W_x(z)^2 +{1\over 2\pi i} \oint_{\CC_-}
{d\z\over z-\z}W_x(\z)\Big[ \sum_{x'} A_{xx'}W_x(-\z )-\sum_n
V'_x(\z)\Big]\Big\rangle_{_{\rm MM}}=0.}
The integration contour $\CC_-$ separates the eigenvalue contour $\CC$
from its reflection image $\bar \CC$ (Fig.  4).  It is also assumed
that the point $z$ is outside the integration contour.

In the Fock space representation, the loop equations are equivalent to
the following set of operator identities satisfied by the right vacuum
$|\Omega\rangle$:
\eqn\opVir{\int _{\CC_-} {d \zeta }\ {: [\p\G_x(\z)]^2 : \over
z-\zeta} \, \, | \Omega \rangle =0+...  \qquad ( x\in\IZ_+ ).}
The omitted terms vanish after multiplying with the left vacuum
$\langle B|$.  In other words for each $x$ the singular at $z=0$ part
of the mode expansion of the energy-momentum tensor
\eqn\defTx{ T_x(z) \equiv \frac{1}{4} :[\p\G_x(z)]^2: } annihilates
the state $|\Omega\rangle$.

\noindent {\it Proof:}\ \ The operator $(\p\G_x)^2$ acts to the right
vacuum as
   \eqn\OPEEO{\eqalign{ T_x(z) |\O\rangle&=\hf \p \G_x (z) \int_{\CC}
   d\z \( 2 { E_x(\z )\over z-\z} - \sum_{x'} A_{xx'} { E_{x'}(\z
   )\over z+\z} \) \Big| \O\Big\rangle\cr &= \int_{\CC}d\z {\p\over
   \p\z} { E_x(\z )\over z-\z} -\sum_{x'} A_{xx'} \int_{\CC}d\z \ \p
   \G_x (z) { E_{x'}(\z ) \over (z+\z) }\Big| \O\Big\rangle }.  }
The first term disappears after multiplying with the left vacuum since
the potential is chosen so that it grows to $+$ infinity at the
endpoints of $\CC$.  The second term is regular in the domain inside
$\CC_-$ and therefore does not contribute to the integral \opVir.

\smallskip
 \bigskip

\noindent \rb {\it Analytical properties of the collective field}

 \medskip

Arguing as in \Idis, we expect that in the quasiclassical limit
$\gs\to 0$ the eigenvalues of the matrix ${\bf M}_x$ condense on some
interval $[-\Lambda, -M]$ on the negative real axis.  The endpoints of
the interval depend on the couplings $T $ and $\k$.  Up to
exponentially small terms the contour $\CC$ can be restricted to the
negative real axis, and the contour $\bar \CC$ -- to the positive real
axis.  Then the contour $\CC_-$ goes from $-\infty$ to $0$ below the
real axis and then back to $-\infty$ above the real axis (Fig. 5b).

The classical value of the resolvent  
\eqn\Wdefz{ W_x^c(z)= {g_s}\< \tr \( z- {\bf M}_{x}\)^{-1}\> }
is a meromorphic function with a branched cut along the eigenvalue
interval and asymptotics at infinity as
\eqn\asyW{ W^c_x(z) \sim {\gs}{N_x \over z }= S_x   \, {\k^2 \over z}.
}
The classical Virasoro constraints \opVir\ imply that $(\p\G_x)^2$ is
analytic in the vicinity of the interval $[-\L, -M]$.  Since $\p\G_x$
is discontinuous across the cut, this means that
   \eqn\dualbc{ \gs\, \p \Ga_x^c(z) = \p V_x(z) -2W_x^c(z) -\sum_{x'}
   A_{xx'} W_{x'}^c(-z) }
satisfies the boundary condition  
\eqn\clasqd{ \eqalign{  \p\G_x^c(z+i0) + \p\G_x^c(z-i0) =0, \quad \ \
z\in [-\Lambda, -M].  } } 
The boundary condition \clasqd, together with \asyW\ and \dualbc,
determines uniquely the meromorphic function $W_x^c(z)$.\foot{Note
that the boundary condition \clasqd\ has a natural interpretation in
the matrix model.  It means that the classical distribution of the
eigenvalues is such that the effective potential $\G_x^c(z)$ is
constant on the eigenvalue interval.  Since the derivative of the
effective potential has a discontinuity across the cut, one has to
take the half-sum of its values on both sides.}

%%%%%%%%%%%%%%%%%%%%%%%%%%%%%%%%%
\epsfxsize=130pt
\vskip 20pt
\hskip 100pt
\epsfbox{ 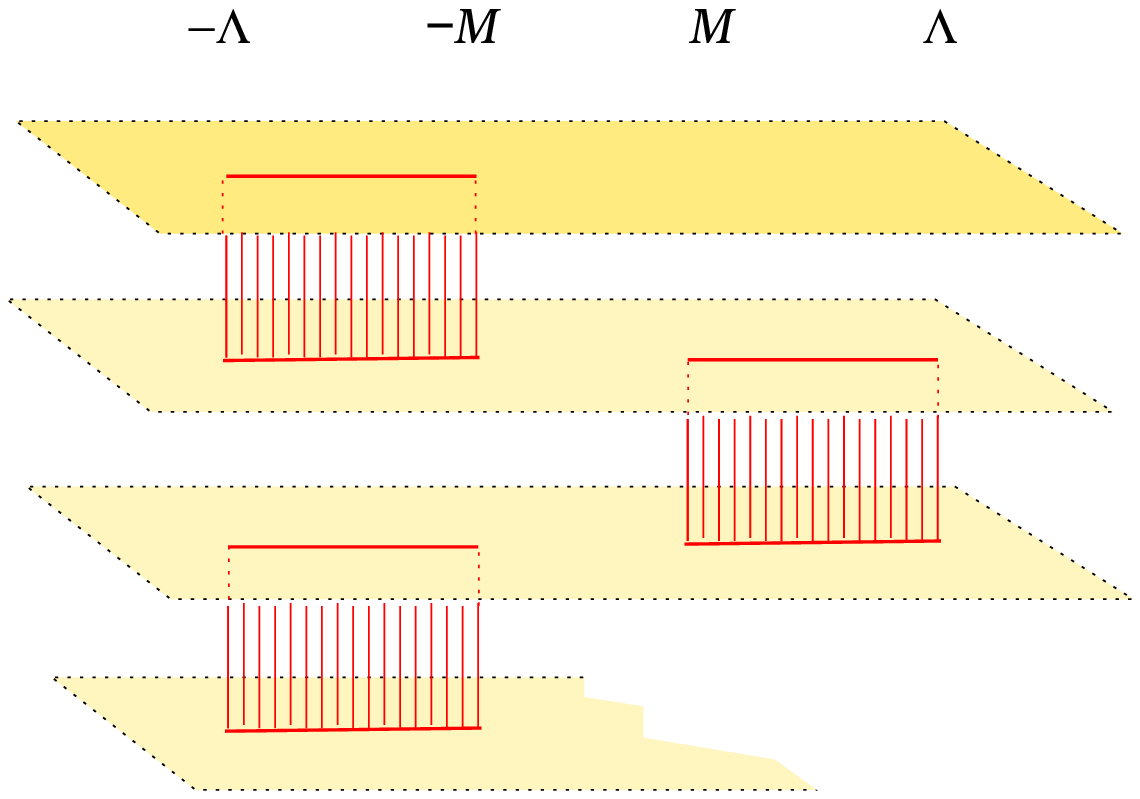   }
\vskip 5pt
%%%%%%%%%%%%%%%%%%%%%%%%%%%%%%%%%%
\centerline{\ninepoint \vbox{\hbox{Fig.  6: Riemann surface of the
classical solution $W^c(z)$.  In the first (physical) sheet }
\hbox{there is only one cut $[-\L, - M]$ while all other sheets
contain two symmetric cuts.} } }
\vskip 10pt
 
\noindent With the particular potential \potv, the classical solution
must be of the form
  \eqn\clascur{ \Phi_x^c(z)={S_x\over S_1} \, \Phi^c(z) , \quad
  \G_x^c(z)={S_x\over S_1} \, \G^c(z) , \quad W_x^c(z) = {S_x \over
  S_1} \, W^c(z) }
%
% \eqn\clsf{ W_x^c(z) = {S_x \over S_1} \, W^c(z) } 
and \clasqd \ becomes a condition for the single function $W^c(z)$:
    \eqn\clasp{ W^c( z +i0) + W^c(z-i0)- 2\cos \pi p_0 \, W^c(-z) -
  \p V(z) =0 , \quad \ \ z\in [-\Lambda, -M].  } 
This is the same equation as in the $O({\rm n})$ matrix model with
${\rm n} = 2\cos(\pi p_0)$ \Ion.  The solution of \clasqd\ for a
general polynomial potential can be expressed in terms of Jacobi theta
functions \EynardNV. Its explicit form in the case of the cubic
potential \potv\ can be found in \Onth .  The Riemann surface of the
function $W^c(z)$ is sketched in Fig.  6.

It is consistent to assume that the quantum field has the same
analytic properties as the classical solution and define its mode
expansion at infinity in terms of a complete set of functions globally
defined on the Riemann surface.  This is the basic assumption for the
iterative procedure for calculation of the $n$-loop functions
perturbatively in $\gs$, known as method of moments, \refs{\ackm,
\EynardNV}.  Our method can be considered as an operator equivalent of
the method of moments.  We thus assume that the boundary condition
\clasqd\ actually holds at {\it operator} level,
\eqn\discEop{\eqalign{
  \p_z[ \Phi_x(z+i0) + \Phi_x(z-i0) - \sum_{x'}
A_{x ,x'} \Phi_{x'} (-z) ]&= \p_z  \[ \G_x(z+i0) + \G_x(z-i0)\]\cr
&=0 
,   \qquad  \qquad  z\in[-\infty, -M],
}
}
and should be understood as a condition on the $n$-point correlation
functions of the collective field $\Phi$.

\subsec{Continuum limit, phase diagram  and  critical points}

\noindent In the ensemble of dynamical triangulations the critical
thermodynamics is controlled both by the temperatore $T$ and the bare
cosmological constant $\k $.  The continuum limit\foot{We do not call
it scaling limit because the theory is scale invariant only at the
critical points.} is achieved, for given $T$, at some critical value
$\k_c(T)$.  The singularity along the line $\k = \k_c(T)$ is due to
the contribution of surfaces of diverging area.  In the continuum
limit, {\it i.e.}, near the critical line $\k = \k_c(T)$, the lattice
can be approximated by a continuous worldsheet and the theory is
described by perturbed Liouville gravity.
  
The behavior of the theory in the continuum limit depends on the value
of $T$.  As in the case of a flat lattice, there are three possible
types of critical behavior, characterized by massive, dilute and dense
loops \Ion \GauK. The critical line $\k = \k _c(T)$ consists of two
branches:
$$\k = \k_c(T)= \cases{\k_c^{I} (T), & if $T>T^*$;\cr &\cr \k_c^{II}
(T), & if $T<T^*$.  } $$
The first branch $\k = \k ^{I} (T)$ describes high-temperature phase
where the area of the graph diverges, while the loops remain finite.
Its equation is given by $\p\k /\p M=0$.  Along this branch the
critical behavior of the partition function is that of pure gravity
with $c_{\rm matter}=0$.  The second branch $\k= \k_{II} (T)$
describes the low-temperature phase of densely packed loops.  It is
determined by $M=0$.  Here the critical behavior is that of a 2D
gravity with matter central charge given by \cden.  The two branches
meet at the critical point $\k =\k ^*,T=T^*$, where both the area of
the graph and the length of the loops diverge.  The double scaling
limit at this point describes the phase of dilute loops with matter
central charge \cdil.

The vicinity of the critical point $T^*,\k^*$ is parametrized by the
renormalized cosmological constant $\mu$ and the temperature coupling
constant $t$, defined as
\eqn\mulam{ \mu \sim \k ^{*} -\k, \qquad t \sim T^*-T. }
We will consider the more general case of finite $t$, where the matter
field is not conformal invariant.  Then the susceptibility $ u =- 4p_0
\p_\mu^2 Z$, where $Z$ is the partition function on the sphere without
boundaries, satisfies the transcendental equation \Onth\foot{The
normalization factor $4p_0$ in the definition of the susceptibility is
fixed by the comparison with the 3-point function, see below.}
\eqn\STREQ{ \mu =\hf u^{1\over p_0} + \frac{1}{ 2-2p_0}\, t u^{ 1-p_0
\over p_0} .  }
This equation describes the flow between the dilute ($t=0$) and the
dense $(t\to\infty$) phases.  The dimension of the coupling $t \sim
\mu^{p_0}$ matches the gravitational dimension $ \delta_{1,3} =
1-p_0$, obtained from $\Delta_{1,3}$ by the KPZ scaling relation
$\Delta = {\delta(\delta+p_0)\over 1+p_0}$ \KPZ. This is consistent
with the conjecture that for finite $t /\mu^{p_0}$ the theory is
described by a perturbation of the critical point $t=0$ by the
Liouville-dressed thermal operator $\Phi_{1,3}$.  Eq.  \STREQ\ is
discussed in the context of perturbed Liouville gravity by Al.
Zamolodchikov \AlZG.

 \subsec{The loop operator and the disk partition function in the
 continuum limit}

\noindent The critical line $\k^{II}_c(T)$ is determined by the
condition that the right edge of the eigenvalue distribution (after
the shift \shiftM) hits the origin while the left end stays at a
finite distance $\L\sim T$.  Therefore in the continuum limit $M\ll
T$.  If we introduce the renormalized length \eqn\Lell{ \ell=
\frac{2}{T} L, } then the loop operator $\tr {\bf M}_x^L$ can be
approximated, after the shift \shiftM, by the exponential
\eqn\loopop{ \tr[({\bf M}_x+\hf T )^L] \approx e^{ {1\over 2} \ell T
\log T } \, \tr (e^{ \ell {\bf M}_x}) .  }
Therefore up to trivial multiplicative factor the operator creating a
boundary of length $\ell$ and height $x$ is
\eqn\defWx{ W_x(\ell) =\gs\, \tr (e^{\ell {\bf M}_x}).  }
 The Laplace transform
\eqn\Wdefz{ W_x(z)= \int_0^\infty {d\ell }\, e^{-z\ell} \, W_x(\ell) =
{\gs} \, \tr ( {1\over z- {\bf M}_{x}}) }
is the operator creating a boundary with a marked point with height
$x$ and (in general complex) boundary cosmological constant $\mu_B=z$.
The $n$-loop amplitude \partfZn\ as a function of the boundary heights
$x_i$ and the boundary cosmological constants $z_i$ is given by
   \eqn\partZz{ Z (x_1,z_1;\dots;x_n, z_n )= \lim _{\gs\to 0}
   \gs^{2-n} \<\!\!\!\!\< \prod _{i=1}^n \tr \log \( z_i - {\bf
   M}_{x_i}\) \>\!\!\!\!\> \, .  }

\newsec{ The collective field theory as a CFT on a Riemann surface}

\subsec{The classical solution in the scaling limit}

\noindent In the following sections we will show how Virasoro
conditions \opVir\ can be solved directly in the continuum limit
$\L\to\infty$.  The solution is completely determined by the classical
value of the resolvent, whose explicit form we give below, and the
target space $\IX$ of the height model.  From now on we will
concentrate on the SRSOS model, 
$$\IX = \IZ_+, \ \ S^{(p)}_x = \sin
(\pi p x), \ \ S_x = \sin (\pi p_0 x), \ \ \ \psi_p(x) = {\sin (\pi p
x)\over \sin (\pi p_0 x)} .
$$
We will give an operator solution of the Virasoro conditions \opVir.
The $n$-loop amplitudes will be expressed as Fock space expectation
values as in \expF, but the left and right vacuum stated $\langle B|$
and $|\Omega\rangle$ will be given a new realization in terms of a
chiral bosonic field on this Riemann surface.  The basic idea of this
approach is that the collective field can be constructed
perturbatively in the string coupling $\gs$ as a bosonic field defined
on the Riemann surface of the classical solution \refs{\Iconfm,
\IHouches}.  The operator solution yields a diagram technique similar
to the one derived for the ADE matrix models \Higkos .

In the continuum limit the classical solution is written in terms of
the uniformization parameter $\t$ by \Onth
\eqn\xoft{\eqalign{ z\ \ \ =&\ \ \ M\, \cosh\t, \cr \p_{z} \Phi^c|_\mu
=&- {1\over 2 \gs } \( {M^{1+p_0} \over 1+p_0} { \cosh(1+p_0) \t } \ -
t \, {M^{1-p_0} \over 1-p_0} {\cosh (1-p_0)\t } \) , \cr \p_\mu\Phi^c
|_{z} =&\ {1\over 2\gs } \, {M^{p_0}\over p_0 }\, \cosh \, p_0 \t .} }
The expressions for the two derivatives are compatible if they lead to
the same expression for the second derivative.  From here we find,
using the formulas
$$\p_\mu = \p_\mu M \( \p_M - \frac{1}{M\tanh\t}\p_\t\),
\ \ \p_z = \frac{1}{M\sinh\t} \p_\t$$
 a  transcendental equation for the modulus $M$:
 \eqn\bentr{ 2
 %{\y\over \x}
 \mu =
  M^2 + \frac{1}{1-p_0} \, t\, M^{2-2p_0}.  
  }
It can be shown \Onth\ that the modular parameter $M$ is related to
the string susceptibility $u$ by
\eqn\defchi{ u = %{1\over p_0}
 M^{2p_0}, } which relation implies the equation of state \STREQ.

From the solution \xoft\  we obtain for the classical effective 
potential  
\eqn\ccurveb{\eqalign{
\p_z\G^{c} (\tau\pm i\pi ) & =\pm  i\,  \p_z\vp ^c(\t), } } where
\eqn\defvpc{
\eqalign{ \p_z\vp ^c(\t):&=2 \sin (\pi x \p_\t )\, \p_z
\Phi^c(\t ) \cr & = 
\frac{\sin(\pi p_0) }{  \gs } \( \frac{M^{ 1+p_0}}{ 1+p_0}\sinh (1+p_0)\t
+t \frac{M^{1-p_0}}{ 1-p_0}
 \sinh (1-p_0)\t \) .  }
}
This function has two cuts, $[-\infty, - M]$ and $[M,\infty]$.  The
two signs in \ccurveb\ correspond to the value of $\p\G^c$ above and
below the left cut.

The field $\vp^c$ has vanishing imaginary part along the interval
$z>M$, which implies that $\p\G^c _x(z)$ has vanishing real part along
its left cut $z<-M$.  Therefore it is expanded at the branch point
$z=-M$ as a series of the half-integer powers of $z+M$.  In the
following we will need the explicit formula for this expansion,
 \eqn\muka{  \p_z\G^c _x(z)= - {1\over M\sqrt{2}\gs} \ { S_x } \sum
 _{n\ge 1} {\mu_{n}\over (2n-1)!!} \(1+{z\over M}\)^{n-1/2}\, . }
The coefficients $\mu_n, \ n=1,2,...  $, which we call moments of the
classical solution, are given by \foot{ The expansion coefficients can
be obtained from the standard representation of the hypergeometric
function (we use that $\sqrt{z+M\over 2M}= i \sinh {\t-i \pi\over
2}$):
$$\eqalign{ \frac{i \sinh (g (\t -i\pi )) }{2g} = \sqrt{\frac{z+M}{ 2
M}}\ \ {}_{2}F_{1}(\hf +g,\hf -g, \frac{3}{2};
\frac{z+M}{2 M})   
= \frac{1}{ \sqrt{2}} \sum_{n\ge 1}
\frac{F_{n-1}(g)}{ (2n-1)!!} (1+\frac{z}{ M})^{n-1/2}\, . }
$$
}
 \eqn\mukag{ \eqalign{
    \mu_n& = 2
    \({M^{ 2+p_0} F_{n-1}(1+p_0)
 + t\, M^{ 2-p_0} F_{n-1}(1-p_0) }\) ,  
 }
  }
where 
\eqn\defFn{ F_n(g) := {(\hf +g)_{n} (\hf -g)_{n}\over n!  }}
and $(y)_n $ is the Pohgammer symbol.  In the particular case $p_0 =
\frac{1}{2}$ (pure gravity) the series \muka\ truncates to the first
two terms.

\subsec{The collective field as an operator field on a Riemann
surface}

\noindent In the parametrization $z=M\cosh\t$ the operator boundary
condition \discEop \ takes the form of a difference equation with
respect to $\t$.  Considered as a function of the uniformization
variable $\t$, the operator field $\Phi_x(\t)\equiv
\Phi_x(z)|_{z=M\cosh\t}$ must

\smallskip

$i)$ be entire function of $\t$,

$ii)$ satisfy the orbifold condition
  \eqn\orbif{
\Phi_x(\t) = \Phi_x(-\t), }

$iii)$ satisfy the discrete Laplace equation in the $\{x,\t\}$ space:
\eqn\disL{ \eqalign{\sum_{x'}A_{x,x'} \Phi_{x'}(\t) & =
\Phi_x(\t+i\pi)+\Phi_x(\t-i\pi). } } 

The last condition can be conveniently written as a difference equation
\eqn\disOLx{\( \cos \pi \p_\t - \cosh \p_x\) \Phi_x(\t) = 0, }
if the restriction to positive heights is imposed by requiring
antisymmetry $\Phi_x= - \Phi_{-x}$.

The condition \disOLx\ means that the components $\{\Phi_x(z)\}_{x\in
\IZ_+}$ and $\{\G_x(z)\}_{x\in \IZ_+}$ of the collective field can be
obtained by analytical continuation from a {\it single} holomorphic
field $\Phi(\t):=\Phi_1(\tau)\,.$ Indeed, \disL\ is equivalent to the
defining functional relation of the Chebyshev polynomials of second
kind
\eqn\PhiaP{\eqalign{ \Phi_x(\t)&=U_{x-1}(\cos \pi \p_{\t}) \
\Phi_1(\tau) \cr &= {\sin\pi x\p_\t \over \sin\pi \p_\t} \
\Phi(\t)\,.} }
For the field $\G_x(\t)$ we have, 
\eqn\PhiD{\eqalign{ \G_x^{}(\t \pm i\pi ) &= 2\Phi_x(\t\pm i\pi) -
\Phi_{x+1}(\t ) - \Phi_{x-1} (\t )\cr
&= \pm 2i \sin \pi \p_{\t} \ \Phi_x(\t) %\cr &
= \pm 2i \sin \pi x\p_{\t}\ \Phi(\t)\cr &=\pm i {\sin\pi x\p_\t \over
\sin\pi \p_\t} \ \vp(\t) } } for real $\t$.  Here we introduced for
convenience a second field $\vp$ related to $\Phi$ by
\eqn\defvp{ \vp(\t) = 2\sin(\pi\p_\t) \, \Phi(\t).  }
Note that $\Phi$ is even, $\Phi(\t) = \Phi(-\t)$, while $\vp$ is odd,
$\vp(\t) = -\vp(-\t)$.

\subsec{Mode expansion and two-point correlator}  

\noindent
We write the operator field $\Phi $ as a sum of positive and negative
frequency parts,
\eqn\inftyexp{ \Phi = \Phi ^+ + \Phi ^- ,}
so that $\Phi^- $ as a function of $z$ is analytic on the Riemann
surface with the point $z=\infty$ removed, and $\Phi^+$ is analytic in
the vicinity of the point $z=\infty$.  The two pieces are given, as
functions of $\t$, by the spectral integrals
  \eqn\mdexp{\eqalign{ \Phi ^-(\t) & = \int_0^\infty \frac{d\nu}{ \nu}
  \ \cosh\nu\t \ \Phi _\nu^-, \cr \Phi ^+ (\t) &=- \int_0^\infty
  \frac{d\nu}{ \nu} \ e^{-\nu\t } \ \Phi _\nu^+,\qquad \Re (\t) >0.  }
  }
Similarly, for the quantum field $\vp(\t)$ we find, using \defvp, the
mode expansion
  \eqn\mdexpvp{ \eqalign{ \vp^-(\t) & = \int_0^\infty d\nu\
  \frac{2\sin \pi\nu}{ \nu} \ \sin\nu\t \ \Phi_\nu^- , \cr \vp ^+ (\t)
  &=- \int_0^\infty d\nu \ \frac{2\sinh \pi\nu}{ \nu} \ e^{-\nu\t } \
  \Phi _\nu^+,\qquad \Re (\t) >0.  } }
The operator amplitudes $\Phi^\pm_\nu$ are assumed to satisfy the
canonical commutation relations
\eqn\Cccr{ [\Phi _\nu^+, \Phi _{\nu'}^- ] = \nu \, \delta(\nu-\nu') }
and the left and right vacuum states are defined by \eqn\vacinf{
\langle 0_{\infty} | \Phi ^- =0, \ \ \ \Phi ^+ | 0_{\infty} \rangle =
0.  }

With this definition of the left and right vacuum states the field has
zero expectation value.  The expectation value \xoft\ can be generated
by replacing the left vacuum by an appropriate coherent state.  The
two-point correlator is given, for $ \Re\t> \Re \t'$, by\foot{ The
normalization of the field was fixed by the condition that short
distance behavior at $\t=\t'$ is compatible with \ccrZ\ and \defPh.  }
\eqn\prphi{\eqalign{ \<\oi | \Phi (\t) \Phi (\t')|\oi \> &=\hf
\log(\t^2 -\t'^2 ) ,\cr \<\oi | \vp (\t) \vp (\t')|\oi \> &=2\sin
\pi\p_\t \sin \pi \p_{\t'} \log(\t^2 -\t'^2 ), } }
where we neglected an infinite constant term.  From here we evaluate
the two-point functions of the original field $\G_x$ and $\Phi_x$:
   \eqn\twpP{ \eqalign{ \< \Phi_x(\t) \Phi_{x'}(\t')\>&= {{\sin(\pi
   x\p_\t ) } \ {\sin(\pi x'\p_{\t'}) }\over {\sin(\pi \p_\t ) } \
   {\sin(\pi \p_{\t'} )} } \ \<\oi |\Phi (\t) \Phi (\t') |\oi\>\cr
   &=\sum_{m=1-x}^{x-1}\, \sum_{m'=1-x'}^{x'-1}\ \<\oi | \Phi(\t+i \pi
   m) \Phi(\t'+i \pi m') |\oi\>\, , } }
\eqn\twpG{ \eqalign{ \< \G _x(\t+i\pi) \G _{x'}(\t'+i\pi)\> =- &
{{\sin(\pi x\p_\t ) } \ {\sin(\pi x'\p_{\t'}) }\over {\sin(\pi \p_\t )
} \ {\sin(\pi \p_{\t'} )} } \ \<\oi | \vp (\t) \vp (\t')|\oi \>\cr &=
-4\, {\sin(\pi x\p_\t ) } \ {\sin(\pi x'\p_{\t} )} \ \<\oi | \Phi (\t)
\Phi (\t')|\oi \> .} }

\bigskip

The two-point correlation functions \twpP\ and \twpG\ are diagonalized
in the momentum space,
\eqn\Phita{ \Phi (\t, p) = \sum_{x\in\IZ_+} \sin \pi p x \,
\Phi_x(\t)\, ,\quad \vp (\t, p) = \pm{1\over i} \sum_{x\in\IZ_+} \sin
\pi p x \, \G_x(\t\pm i\pi)\, .  }
The two fields are related by \defvp, \eqn\vpPhi{ \vp (\t, p) = 2
\sin\pi\p_\t\, \Phi (\t, p) .}

Performing the sum in $x$ in \Phita\ we obtain the mode expansion of
$\Phi(\t,p)$ and $\vp(\t,p)$, which is a restriction of the spectral
integral \mdexp\ to a discrete sum over the momenta $\nu = \pm p \,
({\rm mod} 2)$.  For $\vp(\t,p)$ we find
\eqn\modexpP{\eqalign{ \vp(\t,p) &= \sum_{n\in\IZ} {1\over p+2n} \(
\Phi^-_{|p+2n|} \sinh (|p+2n|\t) +\Phi^+_{|p+2n|} e^{ - |p+2n| \t}\).
} }
The two-point function of $\vp(p,\t)$ is, for $\Re(\t-\t')>0$,
\eqn\serP{ \eqalign{ \<\oi|\vp(p,\t) \vp(p', \t') |\oi\> &=%{1\over 4}
\sum_{n\in{\Bbb Z}} {e^{-|p+2n|(\t-\t') }-e^{-|p+2n| (\t+\t')}\over
|p+2n|}\ \d^{(2)} (p-p') } .}
Here and below $\delta^{(2)}$ denotes the periodic $\delta$-function
\eqn\delltad{ \d^{(2)} (p) :=\sum_{n\in\IZ} \d(p+2n).}
The time-ordered correlator
\eqn\defG{\eqalign{ \ \d^{(2)} (p-p') \ G(\t,\t';p)&= \theta[\Re(\t
-\t')]\, \< \oi|\vp(p,\t)\vp(p', \t')|\oi\> \cr & + \theta[\Re(\t'
-\t)]\, \< \oi|\vp(p',\t')\vp(p, \t)|\oi\> } }
is further diagonalized as
\eqn\idP{ \eqalign{ \int_0^\infty d\t d\t' \sin(E\t) \sin(E'\t')\,
G(\t,\t'; p) &=%{\pi\over 2}
2\pi\, \delta(E-E') \ { G(E,p)},} } \eqn\Gepe{ \eqalign{ G(E,p) & =
\sum_{n\in{\Bbb Z}} \ {1 \over E^2 + (p+2 n)^2} = \ { \pi \sinh \pi E
\over E }\ {1\over 2\cosh \pi E - 2\cos \pi p} .  } }
Finally, using the relations \Phita\ and \vpPhi\ we find the spectral
integral for the the two-point correlator \twpP:
\eqn\Annul{\eqalign{ \< \Phi_{x_1}(\t_1) \Phi_{x_2}(\t_2) \>&= {2\over
\pi}\, \int_0^\infty dE\int _{0}^1dp\ \cos E\t_1\, \cos E\t_2 \, \sin
\pi p x_1\, \sin \pi p x_2\, A (E, p)\cr A(E,p) &= \ { \pi \over E
\sinh \pi E }\ {1\over 2\cosh \pi E - 2\cos \pi p} .  }}

This is the well known expression for the annulus amplitude in
momentum space \refs{\MooreIR, \KostovHN, \Idis }.  The collective
field $\Phi_x(\t)$ creates Cardy type boundary state $|\t, x\rangle$.
The intermediate states in the spectral integral are Ishibashi states
are characterized by two quantum numbers, the Liouville energy $E$ and
matter momentum $p$.

The spectral decomposition \Annul\ is valid also for $ADE$ string
theories, with the integral over momenta replaced by a discrete sum
over Coxeter exponents.  The annulus amplitude for the $A$-series, or
RSOS models coupled to gravity, was derived using the world sheet CFT
approach in \Martinec , see also \KKOPNS\ for an expression analogous
to \twpP.

\newsec{ Operator solution of Virasoro constraints on the Riemann
surface }

\noindent The idea of the operator solution of Virasoro constraints
\opVir\ is that it is possible to introduce another Fock space,
associated with the the mode expansion of the fields $\G_x(z)$ near
the branch point $z=-M$ .  Then the problem reduces in the scaling
limit to the already solved problem of perturbed $c_{\rm matter} =-2$
gravity \DVV, where the operator solution of Virasoro constraints has
been given in terms of a twisted boson.  In this section we first
define the mode expansion of the field $\G_x$ in the half-integer
powers of $z+M$ and the associated bare Fock vacua, $\langle 0_\tw |$
and $| 0_\tw \rangle$ Then we write down the solution of Virasoro
constraints in terms of the corresponding oscillator modes.  This will
lead to the operator replacing the right vacuum $|\O\rangle$.
Finally, we will construct the operator replacing the coherent state
$\langle B|$.  As a result we will obtain a diagram technique similar
to the one obtained in \Higkos\ for the $ADE$ models.

\subsec{Mode expansion at the branch points $z=-M$}

\def\gh{ \hat \gs }

\noindent In the following we will use also the rescaled string
interaction constant and moments defined by
\eqn\mukagh{ \hat \gs = {\gs\over \mu_1}, \qquad \hat \mu_n =
{\mu_n\over\mu_1} .}
For each $x$, the operator field $\G_x(z)$ behaves near the branch
point $z=-M$ as a twisted boson.  Its mode expansion is given by a
series of half-integer powers of $z+M=2 M \cosh^2 {\t\over 2}$, now
positive and negative,
\eqn\alphxa{\eqalign{ \p _z \G_x(z) = & \ \frac{1}{ M\, \sqrt{2} }
\sum_{n\ge 0} (a^{\dagger}_{n,x} - \delta_{n,1} \frac{S_x}{ \gh} )\
{(1+\frac{z}{M})^{n-{1\over 2}} \over (2n-1)!!}\, \cr & +\frac{1}{ M\,
\sqrt{2} } \sum_{n\ge 0} a_{n,x}\, (2n+1)!!\
(1+\frac{z}{M})^{-n-{3\over 2}} .}}
In this expansion we dropped all terms in the expansion of the
classical value \muka\ except the first one, proportional to $
\mu_1$.\foot{If we had dropped the whole classical field, Virasoro
constraints would become singular and would have no solution \DVV.}
The rest of the classical background will be later reintroduced as a
perturbation.

The operators $a_{n,x}$ and $a^\dagger_{n,x}$ and the twisted left and
right Fock vacua satisfy
% %
\eqn\ccra{ [a_{n,x}, a^{\dagger}_{n',x'}]=\delta_{n,n'}\delta_{x,x'}.
}
\eqn\deftw{ \langle 0_\tw| a_{n,x}^{\dagger} = 0, \ \ \ \ a_{n,x}
|0_\tw \rangle=0 \qquad (n\ge 0, \ x\in \IZ_+).  }
The vacuum state $|0_\tw \rangle$ can be thought of as the direct
product of an infinite number of twist operators associated with the
branch points of the Riemann surface \refs{\CMM, \IHouches}.
The two-point function of the field $ \G_x(z)$ in the twisted vacuum
is
\eqn\twistcor{ \langle 0_\tw| \G_x(z) \, \G_{x'}(z') |0_\tw \rangle=
\delta_{x,x'}\, \ln{ \cosh{\t\over 2} - \cosh {\t'\over 2}\over
\cosh{\t\over 2} + \cosh {\t'\over 2}}.  }
The creation operators $a^{\dagger}_{k,x}$ can be expressed by the
generating function
\eqn\bpayy{ \sum_{n=0}^{\infty} {u^n \over n!} (a^{\dagger}_{n, x} -
\d_{n,1} {S_x\over \gh} ) = \sqrt{2} {} \oint\limits_{\CC_-} {dz\over
2\pi i}\ { \p_z \G _x(z) \over \sqrt{ 1+{z\over M} -2u}}.  }
where the contour $ \CC_-$ encircles the cut $[-\infty, -M]$ of
$\G_x(z)$.
Due to the vanishing of the real part of $\G_x$ along this cut the
contour reduces to a closed circle around $-M$.

We will actually need the expansion of the field $\vp(\t, p)$, which
is written in terms of the Fourier transformed creation and
annihilation operators
\eqn\momaa{\eqalign{ a_k(p)&= \sum_{x\in\IZ_+} \sin( \pi p x)\, a_{k,
x} \cr a_k ^{\dagger}(p)&= \sum_{x\in\IZ_+} \sin( \pi p x)\, a_{k, x}
^{\dagger}, } }
where the canonical commutation relations have the form \eqn\CCRm{
[a_n(p), a^{\dagger}_{n'}(p')]=\delta_{n,n'}\, \delta^{(2)}(p-p') .  }
Then the twisted propagator in momentum space is given by replacing
$\delta_{x,x'}\to \delta^{(2)}(p-p') $ in \twistcor.

The formula \bpayy\ can be used to invert the expansion \alphxa\ for
the creating operators, see Appendix B. We have in momentum space the
operator identity
\eqn\akdpaf{\eqalign{
a^{\dagger} _{k}(p)- \d_{k,1}\  \gh^{-1} { \delta^{(2)}(p-p_0) }%
  = \, F_k(\p_\t)\, \p_{\t}\, \vp^\dagger(\t , p )\Big|_{\t=0} }}
where $\vp^\dagger$ is given by the Fourier transform of the first sum
in \alphxa\ and the differential operator $F_k(\p_\t) $ is defined by
\defFn, with $g\to\p_\t$,
$$ F_n(\p_\t) := {(\hf +\p_\t)_{n} (\hf -\p_\t)_{n}\over n!  }.
$$
  We shall also use a weaker version of \akdpaf\ also derived from
  \bpayy\ in which the negative mode part in \inftyexp\ appears in the
  r.h.s.,
\eqn\akdp{ \big(a^{\dagger} _{k}(p) - \d_{k,1}\ \gh^{-1}
\delta^{(2)}(p-p_0) \big) |0_\infty\rangle = F_k(\p_\t) \p_\t
\vp^{-}(\t,p) \Big|_{\t=0} |0_\infty\rangle \,.  }

 \medskip

\subsec{Operator solution of Virasoro constraints  on the Riemann surface}

\noindent The right physical vacuum state will be constructed from the
modes of the $Z_2$-twisted boson.  The conformal invariant physical
vacuum $ |\Omega \rangle $ must be of the form
\eqn\rvs{ |\Omega\rangle= \prod_{x\in \IZ_+} \Omega_x|0_\tw\rangle }
where the operator $\Omega_x$ represents a formal series expansion in
terms of the creation operators $a_{n,x}^{\dagger}$.  The coefficients
of the series are fixed by Virasoro constraints \opVir, where the
stress-energy tensor \defTx\
is expressed in terms of the oscillator modes associated with the
expansion \alphxa,
\eqn\ttem{ T_x(z) = \hf \oo[\p\G_x(z) ]^2\oo= {1\over M^2} \sum_n
L_{n,x}\, (1+\frac{z}{M})^{-n-2}, } where $\oo\ \ \oo$ means the
normal ordering with respect to the mode expansion \alphxa.  We write
down explicitly the first two of the generators $L_n$, $n\ge -1$,
\eqn\Vg{\eqalign{ &L_{-1,x} ={1\over 2 } \(\sum_{m=0}
(a^\dagger_{m+1,x} - \delta_{m,0}\, {{S_x\over\gh} }\, )a_{m,x}
+{1\over 2 }{(a^\dagger_{0,x})}^2\)\,, \cr &L_{0,x} ={1\over 2}
\(\sum_{m=0}(2m+1)(a^\dagger_{m,x} - \delta_{m,1}\, {S_x\over\gh}
)a_{m,x} \)+ {1\over 16 } \, .  }}

We are looking for a perturbative in $\hat\gs$ solution of the
conditions of conformal invariance
$$
L_{n,x} \, | \O \rangle=0\,, \qquad n\ge -1\,,
$$
of the right physical vacuum state \rvs.  The problem is identical
(for each $x$) to the one of solving Virasoro constraints in pure
gravity.  The spectrum of operators in this theory is described by the
flows of the KdV hierarchy \refs{\DVV, \WittenKon} .  An important
property of the theory is that there is a conserved charge (Òghost
numberÓ) with a distinct background charge for each genus, so a
specific correlation function can be non-zero for at most one genus.
The explicit form of the operator $\O_{x} $ is
\eqn\FmomO{ \O_x =\( S_x\over\gh \)^{-{1\over 24}} \!\!\!\!\!  \exp\(
\sum_{n, g \ge 0 } \({ S_x\over \gh }\)^{2-2g-n} \hskip -0.8cm \sum_{
k_1+...  +k_n = n+3g-3} \hskip -0.4cm { \< \t_{k_1}\cdots \t_{k_n}\>
_g \over n!  } \ a^{\dagger} _{k_1,x}\cdots a^{\dagger}_{k_n,x} \) }
where $ \< \t_{k_1}\cdots \t_{k_n}\> _g\ $ is the genus $g$
correlation function in the topological $(c_{\rm matter}=-2)$ gravity.

{\it Note:} The generating function of these correlation functions can
be written as
\eqn\lopf{\eqalign{
Z_\tw(t): &= \ \langle 0_\tw| e^{ %{1\over \gh}
 \sum_{k\ge 0} t_{k, x} a_{k,x}} \O_{x}|0_\tw\rangle\cr &= \({
 S_x\over \hat \gs}\)^{-{1\over 24}}\!\!\!\!  \exp\(\sum_{n, g\ge 0 }
 \({ S_x \over \hat g_s} \)^{2 -2g-n} \hskip -0.8cm \sum_{ k_1+...
 +k_n = n+3g-3} \hskip -0.4cm { \< \t_{k_1}\cdots \t_{k_n}\> _g \over
 n!  } \ t_{k_1}\cdots t_{k_n}\), }} \eqn\wkkk{\eqalign{ \({ S_x\over
 \gh}\)^{2-2g -n} \< \t_{k_1}\cdots \t_{k_n}\>_g &= {\p^n \over \p
 t_{k_1 }...\p t_{k_n}} \ \log \CZ_x(t) _{|_{t=0}}\cr &= \langle
 0_\tw| a_{k_1,x}\dots a_{k_n,x} \O_{x}|0_\tw\rangle_{|_{\rm conn}}\,.
 }} The Virasoro generators are realized as differential operators,
 i.e., $a^\dagger _{k,x} \to t_k\,, $ $a_{k,x}\to \p_{t_k}$.  The
 formulas in \IZk\ differ from \Vg\ and \lopf, \wkkk\ by the rescaling
 $ S_x/\gh\to 1$.

We will restrict ourselves to the genus zero correlation functions and
consider only the piece with $g=0$ in the sum in \FmomO , for which
\refs{\Dijk, \IZk}
\eqn\ssts{ \langle \t _{m_1}...\t _{m_n}\rangle_{_0} = {(m_1+...+m_n)!
\over m_1!...m_n!} , \ \ \ m_1+...+m_n= n-3.  }
\noindent In momentum space we have, neglecting the overall constant,
\eqn\ptApa{ \eqalign{ |\Omega\rangle&= \exp \Big( {\hat \gs}^{n+2g-2}
\sum_{ n\ge 3 } { 1 \over n!} \int \limits _{-1}^1 dp_1...dp_n \hskip
-0.8cm \sum_{k_1+...+k_n= n-3} \hskip -0.8 cm \ V_{k_1...
k_n}(p_1,...,p_n) \, a_{k_1}^{\dagger}(p_1) ....
a_{k_n}^{\dagger}(p_n) \Big) \ |{0_\tw}\rangle }} with
\eqn\vertp{ V_{k_1...  k_n} (p_1,...,p_n) \ =\ \ {(k_1+...+k_n)!\over
k_1!...k_n!} \, N(p_1,...,p_n) , }
\eqn\Nppp{ N_{p_1,...,p_n} = \sum_{x\in \IZ_+} S_x{}^2 \
\psi_{p_1}(x)...\psi_{p_n}(x) .  }

 \subsec{Classical background and two-point correlator in terms of
 twisted bosons}

\noindent
The gaussian field $\p\G_{x} (\t)$ is completely characterized by its
vacuum expectation value \mukag\ and its two-point correlator \twpG.
In terms of the mode expansion \mdexpvp, the left physical vacuum
$\langle B|$ in \FockR\ can be expressed as the coherent state
associated with the vacuum $\langle 0_\infty|$ that generates a vacuum
expectation value.  Now we want to express it in terms of the mode
expansion \alphxa\ and construct $\langle B|$ as a perturbation of the
left twisted vacuum $\langle 0_\tw|$.  In this case the perturbation
is not just a coherent state but rather Bogolyubov transformation,
which changes also the two-point correlator.
\eqn\ptA{\langle B| =\langle 0_\tw |\exp \Big({1\over 2}
 \sum_{k,k' \ge 0} \sum_{x, x'\in \IZ_+} D^{x,x'}_{k,k'} \,
a_{k,x} \, a_{k',x'} -{  S_x\over \gh}
 \sum_{ k\ge 2 } \sum_{x\in \IZ_+} \hat \mu _{k} S_x\, a _{k,x} \Big) .  }
In momentum space with \eqn\Dxp{\d^{(2)}(p-p')\
D_{kn}(p)=\sum_{x\in\IZ_+} \sin (\pi p x)\, \sin (\pi p' x')\,
D_{kn}^{x,x'} } \ptA\ is rewritten as
\eqn\ptAp{
\langle B|=\langle 0_\tw  |\exp%{\over \hat\gs^2}
\Big(  {1\over 2} \sum_{k,k' \ge 0} \   \int\limits_{-1}^1 dp\
D_{k,k'}(p) \  a_{k}(p)  \, a_{k'}(p)
- { 1 \over\gh}
\sum_{ k\ge 2 }  \hat \mu _{k}  \,  a_{k} (p_0)\Big)\,.
}
The left vacuum defined in this way has the properties
\eqn\Btdp{ \langle B| \ (a^{\dagger}_{n,x} - \d_{n,1} {S_x\over\gh})
|0_\tw \rangle =- \hat \mu _{n} {S_x\over \gh}\, , } \eqn\lvb{ \langle
B| \ (a^{\dagger}_{n,x}- \d_{n,1} {S_x\over\gh} ) \
(a^{\dagger}_{n',x'} - \d_{n',1}{S_x\over\gh} S_{x'} )|0_\tw \rangle =
\hat \mu _{n} \hat \mu _{n'} {S_xS_{x'} \over\gh^2}+ \,
D^{x,x'}_{n,n'}\, .  }
 The first of these identities is equivalent to 
 \eqn\aver{ \langle B|
 \G_x (z) |0_\tw \rangle=\langle 0_\infty | \G_x(z) |0_\infty \rangle
 + \G_x^c(z)=\G_x^c(z) }
 The unknown kernel $D^{x,x'}_{k,k'}$ in the second identity \lvb\ is
 determined requiring that \foot{ This identity as well as \aver\ is
 consistent with the qualitative identification of the left vacuum
 $\langle B|$ with $\langle 0_\infty| e^{c \Phi^+_g}$, where the
 positive mode $ \Phi^+_{\nu=g}$ generates the classical part
 $\Phi^c(\t)$ of $\Phi^-(\t)$.  On the other hand we can identify the
 right vacua $|0\tw \rangle$ and $|0_\infty\rangle$.}
\eqn\avertwo{ \langle B| \G_x (z) \G_{x'}(z') |0_\tw \rangle=\langle
0_\infty | \G_x(z) \G_{x'}(z') |0_\infty \rangle + \G_x^c(z)
\G_{x'}^c(z').  }
From  the expansion \alphxa\ we obtain
\eqn\avertwob{ \langle B| \G_x^\dagger (z) \G_{x'}^\dagger(z') |0_\tw
\rangle= \langle 0_\infty | \G_x(z) \G_{x'}(z') |0_\infty \rangle -
\langle 0_\tw | \G_x(z) \G_{x'}(z') |0_\tw \rangle + \G_x^c(z)
\G_{x'}^c(z') }
where the twisted 2-point function is given in
\twistcor.

To calculate the kernel $D_{n n'}(p)$ in momentum space, we apply the
operator identity \akdpaf\ to the Fourier image of \avertwob \ and
compare with the Fourier image of \lvb.  As a result we obtain an
expression for $D_{kk'}(p)$ in terms of the difference of the 2-point
functions in \avertwob.  For the Fourier transform \Dxp\ the relation
reads
\eqn\dakdp{\eqalign{ D_{kk'}(p)= \p_\t\p_{\t'} F_k(\p_\t)
F_{k'}(\p_{\t'}) D(\t,\t',p)|_{\t=0=\t'}\,, } } where
\eqn\CGp{\eqalign{ \d^{(2)} (p-p') \ D(\t,\t'; p) &=%4
 \langle \oi |
\vp (\t,p) \vp (\t',p')| \oi \rangle\cr & - %4
\d^{(2)} (p-p') \ \langle 0_\tw | \vp (\t) \vp (\t') |0_\tw \rangle }
}
It is useful to note that the second term in \CGp\ is equal to the
first term at $p=\hf$:
\eqn\expsG{ %4
\langle 0_\tw | \vp (\t) \vp (\t') |0_\tw \rangle
=%4
G(\t,\t',p=\hf) = - \ln{ \sinh{\t\over 2} - \sinh {\t'\over 2}\over
\sinh{\t\over 2} + \sinh {\t'\over 2}}.  }
The derivative of the first term in \CGp\ computed from \serP\ takes,
up to the $\d$-function factor, the simple form
\eqn\smpl{ %4
\p_\t \langle \oi | \vp (\t,p) \vp (\t',p)| \oi \rangle ={ \cosh
(|p|-1) (\t+\t')\over \sinh (\t+\t')} - { \cosh (|p|-1) (\t-\t')\over
\sinh (\t-\t')}\,.  }
The difference in \CGp\ is smooth everywhere and can be expanded in
Taylor series at the point $\t=\t'=0$:
\eqn\teri{\eqalign{ &\p_\t\, D(\t,\t'; p) =\p_\t\, (|p|-\frac{1}{
2})(|p|-\frac{3}{ 2}) \t\t ' \Big[ 1+\frac{1}{12}
(p^2-2|p|-\frac{3}{4}) (\t^2+\t'^2) \cr &+ \frac{1}{ 3}
\big(p^4-4|p|^3+\frac{5}{4} p^2 +\frac{11}{ 2} |p| +\frac{33}{
16}\big) \big(\frac{1}{ 36} \t^2\t'^2 +\frac{1}{ 120}
(\t^4+\t'^4)\big)+\cdots \Big]\,, } } e.g.,
\eqn\expl{\eqalign{ D_{00}(p)&= (|p|-\frac{1}{ 2})(|p|-\frac{3}{ 2}),
\cr D_{01}(p)&= - \frac{1}{ 2}(|p|-\frac{1}{ 2})(|p|-\frac{3}{ 2})
(|p|+\frac{1}{ 2}) (|p|-\frac{5}{ 2}) ,\cr D_{11}(p)&= \frac{1}{ 3}
(|p|-\frac{1}{ 2})(|p|-\frac{3}{ 2})(|p|+\frac{1}{ 2}) (|p|-\frac{3}{
2}) (p^2 -2p-\frac{9}{ 4}), \ \ \ {\rm etc.} \cr} }

  \subsec{Diagram technique for the   correlation functions}

\noindent Now we are in a position to compute of the $n$-loop
amplitudes.  The scaling limit of the amplitudes \partZz\ is described
as in \WPhi\ by the positive mode part of the field $\Phi(z)$.  Thus
the $n$-point function for $n\ge 3$ will be defined as
\eqn\nptF{ \<\Phi(\t_1,p_1)\dots \Phi(\t_n,p_n)\>_{{\rm genus \ zero}}
=\gs^{2-n} \langle B| \Phi^{+}(\t_1,p_1)\dots
\Phi^{+}(\t_n,p_n)|\O\rangle.  }
The field $\Phi^{+}(\t,p)$ is the Fourier transform of the loop
creation operator which is assumed to annihilate the twisted vacuum
$\Phi^{+}(\t,p)|0_\tw\rangle=0$.

\noindent In order to evaluate the $n$-point amplitudes \nptF \ we
will also need the commutation relation between the loop operator
$\Phi^+(p, \t)$ and the creation operators $a_n^\dagger$.  This can be
done using the linear relation \akdp\ between $a^\dagger_k$ and
$\vp^-(p,\t)$ and the expression for the commutator of $\vp^+$ and
$\vp^-$, which is equal to the two-point correlator \serP. Using the
expression \smpl\ for the $\t$-derivative of the propagator, we find
\eqn\legFa{\eqalign{ \< 0_\infty| \vp(\t, p) a^\dagger_k(p)
|0_\infty\> = 2 \, {(\hf - \p_\t)_k (\hf + \p_\t)_k\over k!} \,
{\cosh(1-|p|)\t\over\sinh\t}.  } }
Then from the relation \vpPhi\ we get
\eqn\legFb{\eqalign{ F_k(p,\t)= \< 0_\infty| \Phi^{+}(\t, p)
a^\dagger_k(p) |0_\infty\>= \, {(\hf - \p_\t)_k (\hf + \p_\t)_k\over
k!} \, {\sinh(1-|p|)\t\over \sin \pi |p| \sinh\t}.  }}

We would like to write the leg factors directly for the local
operators obtained by shrinking the boundaries to punctures.  For this
purpose we expand the loop operator $\Phi^+$ as a power series in $z$.
The allowed powers are of the form $z^{-\nu}$ with $\nu = |p+2n|,\
n\in\IZ$.  The corresponding amplitudes create local operators on the
world sheet with gravitational scaling dimensions $\hf(\nu-p_0)$.
Therefore the leg factors for the local operators are given by the
coefficients in the expansion of the leg factors for loops,
\eqn\powF{ F_k(p,\t) = \sum _{\nu = |p + 2n|, n\in \IZ} F_{k}(\nu) \
(z/M)^{-\nu}.  }
The coefficient corresponding to $\nu = p$ is obtained by replacing
the last factor on the r.h.s. of \legFb\ with $M^p$ and substituting
the derivative $\p_\t^2$ by $\nu^2 =p^2$.  Thus the $n$-tachyon
correlation function $G(p_1,...  , p_n)$ is obtained from the $n$-loop
amplitude \nptF\ by replacing
\eqn\legf{ F_k(\t,p) \ \to   M^{p} \, F_k(p) =    M^{p} \,
 {(\hf -p)_k(\hf+p)_k\over k!} \,
.} 
These are the leg factors for the order operators with $|p|<1$.

Now we are able to evaluate the $n$-point function $G(p_1,...  , p_n)$
of order operators by performing the necessary number of Wick
contractions in \nptF. This amounts to a sum of Feynman diagrams
composed by vertices, propagators, tadpoles and leg factors.  The leg
factors come from the commutators \legFb, while the tadpoles are
associated with the action of $a^\dagger$ on $ \langle B|$ as in
\Btdp.  The tadpoles are proportional to the normalized moments $\hat
\mu_n= \mu_n/\mu_1$ and depend on $t$ and $\mu$ through the
dimensionless coupling
 \eqn\defhatt{ \hat t = {t\over M^{2 p_0}} .  } The first moment
 $\mu_1$, eqn.  \mukag, then reads \eqn\defmuone{ \mu_1 = 2M^{2+p_0}
 (1+\hat t).  }

\noindent Since the string coupling constant enters only through the
ratio $\hat \gs = \gs/\mu_1$, the terms retained in the perturbative
expansion of an $n$-point function are only those with an overall
factor $\mu_1^{n-2}$.  The vertices can be attached either to tadpoles
or to the propagators or to leg factors.  We summarize the Feynman
rules in Fig.  7.

%%%%%%%%%%%%%%%%%%%%%%%%%%%%%%%%%
    \epsfxsize=250pt
   \vskip 20pt
  \centerline{
   \epsfbox{ 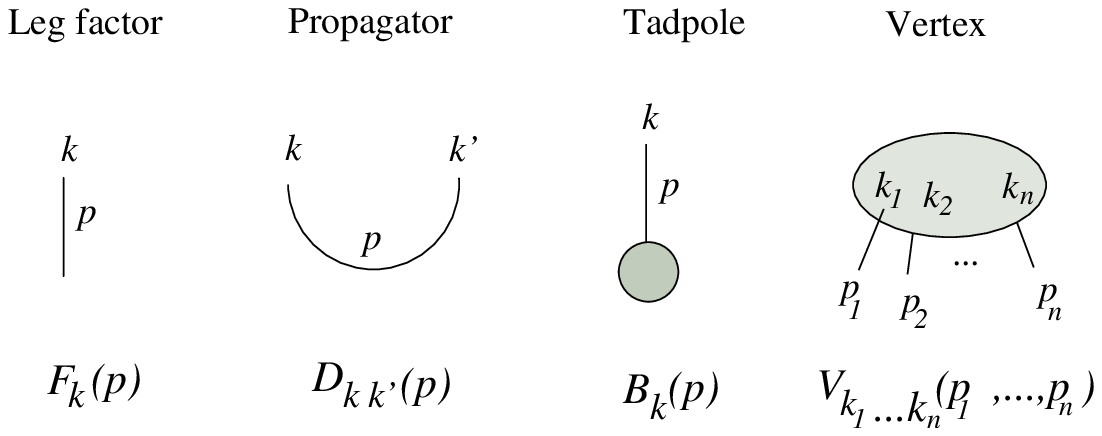 }} \vskip 5pt
%%%%%%%%%%%%%%%%%%%%%%%%%%%%%%%%%%

   \centerline{\ninepoint Fig. 7 : Feynman rules for the correlation
 functions }

    \vskip 10pt

   \bigskip

{\it Feynman rules:}
 
 \bigskip

 \noindent
\rb External line  factors (leg factors)  $   F_k(p)$, where
\eqn\defpi{ \eqalign{ F_0(p)&=1, \ F_1(p)= \frac{1}{4} - p^2, \quad
\dots ,\cr \quad F_k(p)& = {(\hf -p)_k(\hf+p)_k\over k!},\ \dots .  }
}

 \bigskip

\rb Propagator $  D_{k, k'}(p)$:
 \eqn\expl{\eqalign{ D_{00}(p) =& (|p|-\frac{1}{ 2})(|p|-\frac{3}{ 2})
 =-F_1(1-|p|), \cr D_{01}(p)=&-F_2(1-|p|) =D_{10}(p) ,\cr D_{11}(p)= &
 -2F_2(1-|p|)-2F_3(1-|p|) , \ \ \ldots\cr}}

  \bigskip

\rb Tadpole  $B_k (p)$:
 \eqn\tadpls{\eqalign{ B_0 (p)&= B_1(p)=0\,,\cr \ B_n (p) &=- \d^{(2)}
 (p - p_0) \, {\hat \mu_n };\cr \hat \mu_n &= { F_{n-1}(1+p_0) + \hat
 t\, F_{n-1}(1-p_0) \over 1 + \hat t } , \quad n\ge 2 \,.  } }

   \bigskip

\rb Vertices $V_{k_1,...,k_n}(p_1,...,p_n)$:
\eqn\vertics{ V_{k_1,...,k_n}(p_1,...,p_n)= {(k_1...+k_n)!  \over
k_1!...k_n!}\ N_{p_1,...,p_n} .  }

 \bigskip

 \bigskip
 
\rb An overall factor
      \eqn\ovrol{ \hat\gs^{ n-2} \prod_i M^{p_i} = \gs^{n-2} \
      {\mu_1}^{2-n} M^{p_1+...+p_n}.}
 \bigskip
 
We derived the Feynman rules for the case of the SRSOS model, where
the momenta belong to the interval $0<p<1$.  One can derive a similar
set of Feynman rules for the $n$-point functions in the SOS model,
where the the momentum interval is $-1<p<1$.  In both cases the
multiplicities $N_{p_1,...,p_n}$ are given by
\eqn\Npboth{ N_{p_1,...,p_n} = \sum_{x\in \IX} S_x{}^{2-n} \ S
^{(p_1)}_x\cdots S^{(p_n)}_x, }
where $S^{(p)}_x$ is defined by \defSpx.

The multiplicities \Npboth\ are periodic in $p\to p+2$ and, in the
case of the SRSOS model, also antisymmetric in $p\to -p$.  The
propagator $D_{k,k'}(p)$ is the same for both models.  It has the
symmetries $D_{k, k'}(p)= D_{k, k'}(-p)= D_{k, k'}(2\pm p).$ The
propagator and the vertices can be defined outside the interval
$|p|<1$ by periodicity in $p\to p+2$.

 %%%%%%%%%%%%%%%%%%%%%%%%%%%%%%%%% 
 \epsfxsize=120pt
%\vskip 20pt
\hskip 100pt
\epsfbox{ 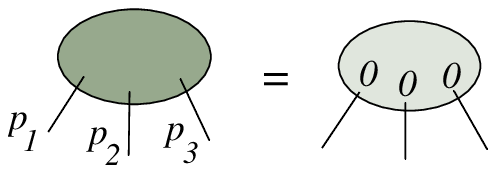   }
\vskip 5pt
%%%%%%%%%%%%%%%%%%%%%%%%%%%%%%%%%%  

\centerline{\ninepoint Fig.  8: The only Feynman diagram for the
three-point function }
 
 \vskip  10pt

The simplest example is the 3-point function (Fig.  8).
The only term of order $g_s$ in the exponent in $|\O\rangle $ which
contributes to this correlator is the one with $\< \t_0\t_0\t_0\>$,
i.e., $k_1=0=k_2=k_3$.  The corresponding leg factors for the local
operators are $F_0(p_i)=1$ and we obtain
\eqn\threep{ G(p_1,p_2,p_3) = {M^{p_1+p_2+p_3}\over \mu_1} 
 V_{k_1,k_2,k_3}(p_1,p_2, p_3)     =
  {M^{p_1+p_2+p_3 -2-p_0}\over  1+ t M^{-2p_0}  } 
{N(p_1,p_2, p_3) }\,.  } 
The 3-point function is proportional to the 3-point fusion
multiplicity.  In the case $p_1=p_2=p_3=p_0$ it is equal to the third
derivative of the partition function on the sphere.  Indeed,
 $$ -\p_\mu^3 Z =  \frac{1}{4p_0} \p _\mu u
 = \frac{  M^{3p_0}}{ \mu_1}.
 $$
This justifies the normalization of the susceptibility, $u = - {4 p_0}
\p_\mu^2 Z= M^{2p_0}$.  To compare with the results from the world
sheet CFT, we should adjust the normalizations of the couplings $\mu$
and $t$, as well as the one of the order operators.

\newsec{ The 4-point function in the SOS and \SRSOS    models}

\subsec{ General formula for the 4-point function}

 \noindent The 4-point function is given by the sum of the three
 Feynman diagrams shown in Fig. 9.  Each Feynman diagram stands for the
 sum of terms that differ by permutations of the external legs.

 %%%%%%%%%%%%%%%%%%%%%%%%%%%%%%%%%
 \epsfxsize=300pt
\vskip 20pt
\centerline{\epsfbox{ 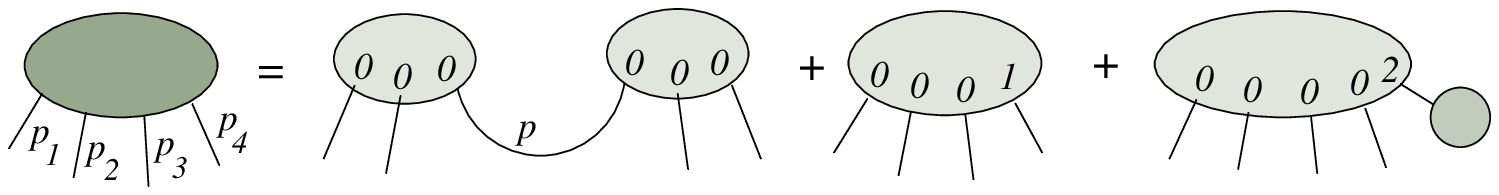 }} \vskip 5pt
%%%%%%%%%%%%%%%%%%%%%%%%%%%%%%%%%%

\centerline{\ninepoint Fig. 9: The diagrams for the 4-point function }

 \vskip 10pt

\noindent We will write the corresponding analytical expression in
such a form that it holds both for the SOS and the SRSOS models.  We
assume that the momenta can have both signs and the momentum
integration is done in the interval $-1<p<1$.  For the SRSOS model the
momentum space interval folds to $(0,1)$ because of the symmetry of
the Feynman rules under $p\to -p$.  We have
\eqn\ivanb{ \eqalign{ G(p_1,p_2,p_3,p_4)&= M^{|p_1|+...+|p_4|}
\mu_1^{-2} \times \cr \times &\Bigg[ \int \limits _{-1}^1 dp \[
F_0(p_1)F _0(p_2) N_{p_1p_2 p} D_{00}(p) N_{-pp_3p_4} F _0(p_3)F
_0(p_4) + {\rm perm.} \] \cr & + \mu_1^{-2} \ N_{p_1p_2p_3p_4}\Big[ F
_0(p_1)F _0(p_2)F _0(p_3)F _1(p_4) + {\rm perm.}\Big] \cr &+
\mu_1^{-2} \ N_{p_0 p_1p_2p_3p_4} \, (- {\hat \mu_2 }) \, F _0(p_1)F
_0(p_2)F _0(p_3)F _0(p_4) \Bigg] } }
Then, using \defpi-\vertics\ we rewrite \ivanb\ as
\eqn\ibis{\eqalign{ &   G(p_1,p_2,p_3,p_4)=  
M^{|p_1|+...+|p_4| }  \mu_1^{-2}  \Bigg(  \Big[ -\hat \mu_2 +
\sum_{s=1}^4\(\frac{1}{ 4}- p_s^2\) \Big]N_{p_1,p_2,p_3,p_4} \cr &+
  \int \limits _{-1}^1 dp \( N_{p_1,p_2,p}N_{-p,p_3,p_4}
+N_{p_1,p_3,p}N_{-p,p_2,p_4}+N_{p_1,p_4,p} N_{-p ,p_2,p_3}\)
\(|p|-\hf\)\(|p|-\frac{3}{ 2}\) \Bigg). }}
This is the general formula for the 4-point function in the continuum
limit.  The 4-point function depends on the scale parameter $\hat t =
t/[M(t,\mu)]^{2p_0}$.  It can be compared with the CFT results at the
critical points $\hat t=0$ and $\hat t\to\infty$, where the world
sheet theory is described by Liouville gravity with matter CFT
component with central charges respectively \cdil\ and \cden.  At the
two critical points $\hat \mu_2$ is given by a numerical constant,
\eqn\muhat{\eqalign{- \hat \mu_2 &= {3\over 4} + p_0^2 + 2{1-\hat
t\over 1+\hat t}\, p_0 \to \ \ {3\over 4} + p_0^2 \pm 2 p_0, }}
while $\mu_1$ scales as a power of $M(\mu, t)$:  
\eqn\muone{ \mu_1\ \sim M^{2\pm p_0 }.  }
Here the sign is $+$ for  $\hat t=0$ (dilute loops)
and $-$ for  $\hat t\to\infty$ (dense loops).

In the following we will write more explicit expressions using the
concrete form of the fusion coefficients in the SOS and SRSOS models.
To simplify the algebraic expressions we introduce an auxiliary
quantity equal to the integral of a function $f(p)$ with respect to
the intermediate momentum $p$ and weighted by the fusion coefficients
involving the four external momenta:
\eqn\sumsp{ \eqalign{& \lll f(p)\rrr\ \ := \int _{-1}^1 dp \, f(p)
\(N_{p_1p_2p} N_{-pp_3p_4} +N_{p_1p_3p} N_{-p_2p_4} + N_{p_1p_4p}
N_{-pp_2p_3p} \).  }}
In particular $\lll 1\rrr = 3N_{p_1p_2p_3p_4}$.  Then the general
formula \ibis\ can be written as
\eqn\ibisf{\eqalign{ G(p_1,p_2,p_3,p_4) &=%\gh^2\,
  {M^{|p_1|+...+|p_4|} \over \mu_1^2} \lll \Big[ \frac{1}{3}\big( -
  \hat \mu_2 + \sum_{s=1}^4\(\frac{1}{ 4}- p_s^2\)\big) +
  \(|p|-\hf\)\(|p|-\frac{3}{ 2}\) \Big] \rrr \cr
&={M^{|p_1|+...+|p_4| -2-2p_0}\over (1+\hat t)^2}\,%\gh^2
\( 2\frac{1-\hat t}{ 1+\hat t}\, p_0+ 4 +p_0^2- \sum_{s=1}^4 p_s^2 +
\lll p^2 - 2|p|\rrr \) }}

\subsec{ The 4-point function in the SOS model}

\noindent
 We first evaluate the 4-point function for a simpler case,
that of the SOS model on random triangulation, whose continuum limit
is supposed to be described by Liouville gravity having a twisted
boson as a matter field.  In this case the wave functions are the
plane waves \orderp\ and momentum-space vertex \Nppp\ is a periodic
delta-function, which describes $u(1)$-type fusion rules:
\eqn\NZmod{\eqalign{ N_{p_1,\dots ,p_n}&= \delta ^{(2)}(\sum_{k=1}^n
(p_k- p_0) + 2 p_0).  }}
Since the fusion coefficients are delta-functions, 
the integral over the intermediate momenta  yields 
$N_{p_1p_2p_3p_3}$ times a sum over the three channels.

We would like to avoid the complications related to the compactness of
the momentum space.  This is why we assume that the external momenta
are sufficiently small, so that the periodic delta functions in
\NZmod\ can be replaced by ordinary ones.  Then we can easily perform
the integration in \sumsp, which leads to the following simple
expression:
\eqn\sumsp{\eqalign{\lll p^2\rrr &= N_{p_1p_2p_3p_4} \big( (p_1+p_2 -
p_0)^2 +{\rm permutations} \Big)\cr &= N_{p_1p_2p_3p_4} \ \big(
\sum_{s=1}^4 p_s^2 - p_0^2 \Big) \,.  } }
(For arbitrary external momenta we would obtain a more complicated
expression, which is periodic in $p_s$ and therefore not analytic.)
Substituting this in \ibisf\ we get the final formula, in which the
quadratic in the external momenta terms cancel:
\eqn\ibisQ{\eqalign{& G(p_1,p_2,p_3,p_4)= {M^{|p_1|+...+|p_4|
-2-2p_0}\over (1+\hat t)^2}\, { N_{p_1p_2p_3p_4} } \times\cr &\times
\Big[ 2 + \frac{1-\hat t }{ 1+ \hat t}\ p_0 - |p_1+p_2-p_0| \
-|p_1+p_3-p_0| -|p_1+p_4-p_0| \Big] .} }
At the critical point $t=0$ this expression coincides with the well
known formula for the 4-point correlator of gaussian matter field
\DiK. Indeed, using the map \chihash\ we write the expression in the
brackets as
\eqn\fourpDK{\eqalign{ Q- |P_1+P_2 -e_0| -|P_1+P_3 -e_0|-|P_1+P_4
-e_0|\, }} with \eqn\PDIL{ P_s = {b p_s }, \ \ Q= b(2+p_0 )= {1\over
b}+b,\ \ e_0 = {b p_0 } = {1\over b} - b; \ \ b = {1\over
\sqrt{1+p_0}} \qquad {\rm} \ \ (t=0).  } The world sheet effective
action for the point $t=0$ is that of Liouville gravity with gaussian
matter:
\eqn\efact{ \eqalign{ \CA & = {1\over 4 \pi}\int d^2 \s \[ (\p_a
\phi)^2 + (\p_a \chi)^2+ (Q \phi + i e_0\chi ) \hat R\,\sqrt{\hat g} +
\l_{_L} e^{ 2b\phi } \] + {\rm ghosts}, } } 
where $ \l_{_L}\sim \mu$.  Therefore, at the critical point $t=0$,
\ibisQ\ reproduces the CFT result of \DiK\ for gaussian matter field.
In the dense phase, $t\to\infty$, the formula \ibisQ\ can be again
interpreted as the 4-point function with gaussian matter, but with a
smaller central charge \cden.  In this case we identify
 \eqn\PDENS{
 P_s =-{p_s\, b'}, \ \ Q= (2-p_0)  b' =  {1\over b'}+b',\
 \ e_0 = -{p_0\, b'} = {1\over b'} -b';
 \  \ b' = { 1\over \sqrt{1-p_0}} 
 \qquad {\rm} \ \ ( t\to\infty).
 }  
At finite $t$ the world sheet theory does not factorize to a matter
and Liouville components.  It can be considered as a perturbation by
the Liouville-dressed thermal operator $\Phi_{1,3}$ of the CFT
describing the dilute phase.\foot{This perturbation is expected to
lead to the periodicity of the correlation functions under $P\to P+2b$
which matches the periodicity $p\to p+2$ in the microscopic theory.
We thank Al.  Zamolodchikov for a discussion on this point.}

\subsec{Four-point function in the \SRSOS model: generic momenta }

\noindent Now we will consider the case of generic momenta in the
\SRSOS model.  We assume again that the moments are sufficiently small
so that the compactness of the momentum space is not felt.  For
generic (but sufficiently small) momentum $p>0$ the wave functions
  \eqn\orderpD{\psi_{p}(x) = { \sin(\pi p x) \over \sin(\pi p_0 x)} }
 lead to a formal expression for the vertices \vertics, which does not
 seem to have a rigorous meaning, even as a distribution.  It is
 therefore difficult to give a meaning of the general formula \ivanb\
 for the 4-point function for generic momenta.  We can however modify
 the definition of the observables to obtain a sensible formula for
 the 4-point function.  Namely we consider the wave function for given
 momentum as a sum of two terms
\eqn\orderch{ \psi_p(x) = \psi_p^{(+)}(x) - \psi_p^{(-)}(x), \qquad
\psi_p^{(\pm )} (x) = {e^{\pm i p x} \over e^{i\pi p_0 x}- e^{-i\pi
p_0 x}}.  }
The two terms, $\psi_p^{(+)}$ and $\psi_p^{(-)}$, are analogs of the
tachyons of positive and negative chirality considered in the
world-sheet CFT analysis \Bulkcft, and are related by $p\to -p$.  With
this replacement we pick up the analytic expression for the 4-point
function valid in the infinitesimal vicinity of $p_1= \dots = p_4 =
p_0=0$.

In the case of irrational $p_0$ we can evaluate the ``multiplicities''
$N_{p_1,\dots, p_n}$ by formally expanding the denominator in
\orderch.  Instead of specifying the chiralities, we will consider
positive as well as negative values of the momenta $p_s$.  Then we can
write the fusion coefficients as  a sum   
 of $\d$-functions, {\it e.g.}
 \eqn\NAtr{\eqalign{ N_{p_1p_2p_3 }&\to \sum_{n=0}^\infty \ \delta \(
 p_1+p_2+p_3 -(2n+1) p_0\) } } \eqn\NAfo{ N_{p_1p_2p_3 p_4}\to
 \sum\limits_{m\ge 0} (m+1)\ \delta\(p_1+p_2+p_3+p_4 - 2(m+1) p_0\), }
which we substitute in the general formula \ibisf.  To write the final
expression we must first evaluate the integral
\eqn\sumsp{\eqalign{& \lll p^2\rrr = \sum_{ n\ge 1} \delta \big(\sum_i
p_i - 2np_0\Big) \times \sum_{k=0}^{n-1} \big( (p_1+p_2 - (2k+1)
p_0)^2 +{\rm permutations} \Big)\,.  } }
The expression multiplying the delta function simplifies to
\eqn\summmm{\eqalign{ \sum_{k=0}^{n-1} \Big(\sum_{j=2}^4 (p_1+p_j)-
(2k+1)p_0\Big)^2 &= n \big(\sum_{i=1}^4 p_i^2 - p_0^2 \big) } }
where we used the identity (assuming $\sum p_s= 2n p_0$)
\eqn\idtity{ \sum_{s=2}^4 (p_1+p_s-np_0)^2 = \sum_{s=1}^4 p_s^2 -
n^2p_0^2.  }
Now we can write the explicit expression for the r.h.s. of \ibisf,
which is again linear in the momenta,
\eqn\ibbis{\eqalign{ G(p_1,p_2,p_3,p_4)=& {M^{p_1+...+p_4 -2-2p_0}
\over (1+\hat t)^2}\, \sum_{n=1}^\infty \delta(p_{\rm tot}-2n
p_0)\times\cr &\times \Big[ n\(2 + \frac{1-\hat t }{ 1+ \hat t}\ p_0
\) - \sum_{k=0}^{n-1} \sum_{s=2}^{4} |p_1+p_s- (2k+1)p_0| \Big]\, , }
}
where $p_{\rm tot} =p_1+p_2 +p_3+p_4$ is the total momentum.  Passing
to world sheet CFT notations, we write \ibbis\ at the critical points
as
\eqn\ibbiPs{\eqalign{& G(P_1,P_2,P_3,P_4)\sim
    \sum_{n=1}^\infty \, \delta\(\sum_1^4 P_i -2n e_0\)%\times\cr&
   \Big[ n Q - \sum_{k=0}^{n-1} \sum_{s=2}^{4} |P_1+P_s- (2k+1)e_0|
   \Big]\,.  } }
This expression is identical with the expressions for the 4-point
tachyon correlation function obtained for the ``diagonal
perturbation'' of the non-rational Liouville gravity \Bulkcft.  More
precisely \ibbiPs\ corresponds to the physical, symmetric in the four
momenta correlators, extracted from the fixed chirality solutions of
the functional relations.  The intermediate states in \ibbiPs\ are
spaced by $2e_0$ and can be interpreted in terms of insertions of a
screening vertex operator (tachyon) with matter component $e^{2ie_0
\chi}$.  This amounts to adding an additional ``diagonal
perturbation'' to the Liouville term:
 \eqn\diag{\CA_{\rm int} ^{t=0}= \int d^2 \s \(\l_{_L} e^{ 2b\phi } +
 \l_{_D} e^{ 2ie_0 \chi} e^{ {2b}\phi } \) .  }
In \Bulkcft\ the functional equations for the correlation functions
were obtained by adding the action obtained by Liouville reflection,
$\tilde \CA_{\rm int} ^{t=0} =\tilde \l_{_L} e^{ 2\phi /b} + \tilde
\l_{_D} e^{ 2ie_0 \chi} e^{ {2\over b}\phi } $.

\subsec{The four-point function of degenerate fields in the \SRSOS
model}

\noindent Here we assumed that $p_i = m_i p_0$ are sufficiently small,
so that the intermediate momenta are smaller than 1.  The wave
functions \orderpD, restricted to the discrete spectrum $p\in p_0
\IZ_+$,
\eqn\cha{ \psi_m (x) = {S^{(p)}_x\over S_x}= {\sin (\pi p_0 m x)\over
\sin (\pi p_0 x)}\,, \qquad p = m p_0, \ \ \ m\in {\Bbb N}\, , }
 form closed Pasquier-Verlinde-like fusion algebra: \eqn\uedno{
 \psi_{mp_0}\psi_{np_0} =\sum _{k=1}^{\frac{1}{2}(m+n-|m-n|)}
 \psi_{(m+n+1- 2k)p_0}\, .  } This is the algebra of the order
 operators.  The multiplication rule in \uedno\ coincides with the
 tensor product decomposition multiplicity of $sl(2)$ irreps of
 dimension $m_i$
 \eqn\sltwo{\eqalign{ &N_{m_1,m_2,m_3}= \cases{ 1 & if $\matrix{
 &|m_1-m_2|+1\le m_3\le m_1+m_2-1 \cr & {\rm and}\ \ m_1+m_2+m_3= {\rm
 odd} }$\ ;\cr &\cr 0 & otherwise} .  \cr }} The 4-point multiplicity
 is correspondingly
 \eqn\fpo{\eqalign{ N_{m_1 m_2 m_3 m_4}&=\sum_{m>0} N_{m_1 m_2 m}\
 N_{m m_3 m_4}\cr &=\hf \(\min(m_1+m_2, m_3+m_4)-\max(|m_1-m_2|,
 |m_3-m_4|\).  }}
The $n$-point multiplicities are given by the general formula \Nppp.
The sum over $x\in \IZ_+$ can be replaced, up to an infinite volume
factor, by an integral w.r. to the compact variable $\theta = \pi p_0
x$:
 \eqn\Nint{ N_{m_1,..., m_n}=\frac{1}{\pi} \int _0^{2\pi} d\theta
 \sin^2 \theta\ \prod_{i=1}^n \frac{\sin(m_i\theta)}{\sin\theta} .  }
To evaluate the sum in \ibis, we first rewrite the expression as
\eqn\ibisa{\eqalign{ & G(m_1,m_2,m_3,m_4) = \lll { -\hat \mu_2
+\frac{1}{4} +3 -p_0^2 \sum_{s=1}^4 m_s^2 + p_0^2m^2 - 2p_0 m \over
\mu_1^2} \rrr } }
where the symbol $\lll f(m)\rrr$ is defined similarly as in \sumsp:
\eqn\sumsm{ \eqalign{& \lll f(m)\rrr\ \ := \sum_{m} \, f(m)
\(N_{m_1m_2m} N_{mm_3m_4} +N_{m_1m_3m} N_{m_2m_4} + N_{m_1m_4m}
N_{mm_2m_3m} \).}}

It is straightforward to check that
\eqn\linrell{\eqalign{ \lll m\rrr  &=N_{m_1m_2m_3m_4}(
\sum_{s=1}^4 m_s - N_{m_1m_2m_3m_4} ) \cr \lll
m^2\rrr  &=N_{m_1m_2m_3m_4}( \sum_{i=1}^4 m_i ^2 \ -
1)\,.  }}
Inserting this  in \ibis\ we obtain the following simple
expressions for the 4-point function of order operators. 
\eqn\fpconj{\eqalign{ G_{m_1m_2m_3m_4}&= {2\over \mu_1^2} \Big((2+
\frac{1-\hat t}{1+\hat t}\, p_0) N_{m_1m_2m_3m_4}\, -
\sum_{m=1}[N_{m_1m_2m} \, (m p_0)\, N_{mm_3m_4} + {\rm
perms.}]\Big)\cr &= {2\over \mu_1^2} N_{m_1m_2m_3m_4}\Big( 2+
\frac{1+\hat t}{1-\hat t}p_0 - p_0\sum_{i=1}^4 m_i + p_0
N_{m_1m_2m_3m_4}\Big).  } }
In the two critical points, $\hat t=0$ and $\hat t\to\infty$, the
formula \fpconj\ for the 4-point function can be compared with the
expression obtained in \Bulkcft\ for the CFT with the ``diagonal
perturbation'' \diag\ using the ground ring relations in the Coulomb
gas approach.  We find agreement, up to an overall constant.  In the
case of finite temperature, at the moment there are no world sheet CFT
results to compare with.

\subsec{The 4-point function in ADE string theories}

\noindent At rational values of the background momentum, $p_0=1/h, \
h\in\IZ_+$, the integers $m_i$ are restricted to the range $1\le m_i
\le h-1$ and the $sl(2)$ multiplicities \sltwo, \Nint\ are replaced by
the Verlinde fusion multiplicities, i.e., the upper bound in \sltwo\
becomes $\min(m_1+m_2-1,2h-m_1-m_2-1)$.  The corresponding relations
\linrell\ do not hold true anymore in general, however the initial
quadratic formula \ibis, derived in \refs{\Higkos, \Icar}, survives
and represents the tachyon 4-point function for $p_i=m_i p_0 $.  This
formula extends to the ADE cases.  The 3-point multiplicities
$N_{m_1m_2m_3}$ of the $ADE$ series are real numbers (square roots of
rationals), defining the structure constants of the Pasquier algebra
\Pasq\ associated with any $ADE$ graph, with the numbers $m_i$
corresponding to the set of exponents\foot{These %structure 
constants
coincide with the relative scalar OPE coefficients of the
% local 
$ADE$
theories \PZ. 
}.

The partition functions of the ADE string theories on a torus are
calculated in \Idis.  They are obtained by summing only over the order
fields propagating along one of the circles of the torus.
Nevertheless they reproduce the expressions obtained from the world
sheet CFT \bk , and there are no configurations that are not taken
into account.  This can be explained by the particular choice of the
time slice, which goes along a domain wall.  The torus is thus
obtained by identifying the boundaries of a cylinder with Dirichlet
type boundary conditions.

The fact that the partition functions can be reconstructed only by
taking into account the diagonal fields is also related to the fact
that there is a world sheet CFT, the diagonal theory in \Bulkcft, in
which these fields form closed algebra.  One can give the following
heuristic explanation of that.  In the case of rigid geometry an order
operator at the point $r$ is constructed by cutting out a circular
domain having as a center this point and large compared to the lattice
size.  In general, there will be loops that enter and leave the
domain, crossing its boundary.  Therefore, when shrinking the domain
to a point in the continuum limit, we will create also other operators
as the thermal operators mentioned before.  If we calculate a
correlation function of two such operators, we will reproduce the
standard OPE expansion of the diagonal fields ($r=s$) in CFT, which
will include non-diagonal fields as well.  Now consider the \SRSOS
model on a fluctuating lattice and repeat the procedure.  In such a
``liquid'' lattice the symmetry is stronger and the notion of circular
domain has no sense anymore.  The domain surrounding the point $r$ is
formless; it has a single geometrical characteristics, its area and
perimeter.  We can restrict ourselves to the set of domains that do
not cross any domain walls, i.e. with Dirichlet type boundary.  Then
the definition of the order operator by shrinking the boundary of the
domain is such that in the 4-point function of order operators there
will be only order operators in the intermediate channels.  This is
exactly what happens in the diagonal worldsheet string theory
\Bulkcft.  The definition of the order operators such that they form
closed algebra is possible only in the theory coupled to gravity.

%%%%%%%%%%%%%%%%%%%%%%%%%%%%%%%%
\newsec{Summary and discussion}
%%%%%%%%%%%%%%%%%%%%%%%%%%%%%%%

\noindent The theory we studied here -- the \SRSOS height model on
random triangulations -- is a good candidate for a generic microscopic
realization of non-rational 2D gravity.  At the rational values of the
background momentum $p_0$ it describes the $A$-series of the $ADE$
string theories.  The model is characterized by semi-infinite target
space, which we identified with the $A_\infty$ Dynkin graph.  The
semi-infiniteness of the target space leads to identification of the
positive and negative momenta, which implies a reflection property of
the correlation functions.  It is also possible to construct a model
that generalizes the height models of the $D$ series coupled to
gravity.  The construction of the $D_\infty$ height model is sketched
in Appendix B.

We formulated the SRSOS model as a conformal field theory on a Riemann
surface representing an infinite branched cover of the complex plane.
The points of the Riemann surface label the FZZ branes in the SRSOS
model.
 
The calculation of the correlation functions is straightforward and it
matches with the world sheet CFT predictions.  Our results are
compatible with the conjecture that the \SRSOS model gives a
microscopic realization of the ``diagonal'' CFT introduced in
\Bulkcft, in which the screening charges are the Liouville dressed
identity operator (with momentum $p_0$) and its charge conjugated
(with momentum $-p_0$).

The \SRSOS model gives formulas only in the physical regions of
momenta, for which the tachyons describe local observables on the
world sheet.  The expression is given by a sum of terms associated
with intermediate channels.  Each such term is a polynomial of the
momenta of the particles involved.  These polynomials are higher order
than the polynomials that occur in the iteration procedure of Di
Francesco and Kutasov \DiK .  In the case of non-rational theory there
are relations that allow to lower the degree of the polynomials.
Using them we have transformed the discrete model 4-point functions
into expressions comparable with the CFT solutions of the ring
relations.  This property does not seem to extend to the case of the
minimal matter theory with rational $b^2$ - a case which is especially
difficult to be analyzed in the CFT approach.  Thus we expect that the
4-point functions in this case are given by expressions similar to the
initial quadratic formula \ibis.  The discrete approach provides as
well the analogs of the ADE correlators, which cannot be treated in
the CFT approach, as the bulk ground ring relations, expressing the
fundamental isospin 1/2 fusion relations, do not apply directly to the
non-diagonal $D,E$ cases.

In this paper we restricted ourselves to the correlation functions of
order operators.  In fact our general formula \fpconj\ is the
generating function of the correlators of $4$ order operators and an
arbitrary number of Liouville dressed thermal operators $\Phi_{1,3}$,
which are given by the expansion coefficients of the thermal coupling
$t$.  The meaning of the results for finite temperature will be
discussed in detail elsewhere \ref\KZ{I. Kostov and Al.
Zamolodchikov, in preparation.}.

\bigskip \noindent {\bf Acknowledgments} \smallskip {\ninepoint
\noindent We thank A. Belavin, Al.  Zamolodchikov and J.-B. Zuber for
the interest in this work and for useful comments.  This research is
supported by the European Community through
RTN EUCLID, contract HPRN-CT-2002-00325,
MCRTN ForcesUniverse, contract MRTN-CT-2004-005104,
MCRTN ENRAGE, contract MRTN-CT-2004-005616,
MCRTN ENIGMA, contract MRTN-CT-2004-005652,
and by the Bulgarian National Council for Scientific Research, grant
F-1205/02.  I.K.K. thanks the Institute for Advanced Study, Princeton,
and the Rutgers University for their kind hospitality during part of
this work.  V.B.P. acknowledges the hospitality of Service de Physique
Th\'eorique, CEA-Saclay, and LPTHE, Paris VI. }

\appendix{A}{Loop gas representation of the correlation
 functions}

\subsec{Correlation functions of order operators in the SOS model}

\noindent The order, or vertex, operator in the SOS model is defined
as the effect of inserting the wave function
  \eqn\orderp{\hat \psi_{p}(x) = e^{ i \pi (p_0-p)x} = e^{2i q x}}
at some point of the planar graph.  We will refer to $p$ as a target
space momentum and to $q=\hf(p_0-p)$ as electric charge.  The
correlation functions of the vertex operators \orderp\ can be
formulated as the partition function of the loop gas with some of the
loop weights modified.  The rule is that the weight of a loop
encircling a total charge $q^{\rm in} =\hf(p_0-p^{\rm in} )$ is equal
to
   \eqn\weightlp{ 2\cos(\pi p^{\rm in}) =2\cos[\pi (p_0-2 q^{\rm in}
   )].  }
Since the reference point (the infinite point for the plane) can be
put anywhere, this rule is consistent only if it gives the same result
with $q^{\rm in}$ replaced with the total charge $q^{\rm out }$
outside the loop.  This is guaranteed by the charge conservation
condition $q^{\rm in} +q^{\rm out } = p_0$ or, in terms of momenta,
$p^{\rm in} +p^{\rm out } =0$.

\medskip

\noindent {\it $\bullet$\ One-point function}

\noindent There is only one vertex operator, $\hat \psi_{-p_0}(x) =
e^{2i\pi p_0 x}$, with non-trivial one-point function.  This operator
is related to the identity operator by the charge reflection
\eqn\chref{ q\to p_0-q\qquad {\rm or} \qquad p\to -p.}
 The charge of this operator compensates the total background charge
 $-p_0$ for the sphere.  The correlation function $\langle\hat
 \psi_{-p_0} \rangle$ is equal to the partition function of loop gas
 with weight $2\cos(\pi p_0)$ per loop.

\medskip \noindent {\it $\bullet$\ Two-point function}

\noindent The two points can be connected by a line that intersects
each loop at most once.  The two-point function $\< \hat \psi_p\hat
\psi_{-p}\>$ is obtained by changing the weight of the
non-contractible loops, {\rm i.e.}, those that intersect the line, to
$2\cos\pi p$.

\medskip \noindent {$\bullet$\ $n$-point function\ \ ($n\ge 3$)}

\noindent Consider the $n$-point function \eqn\nptfff{
G_{p_1,...,p_n}= \<\hat \psi_{p_1}\hat \psi_{p_2} \dots \hat
\psi_{p_n}\>, \qquad \sum _i p_i = (n-2)p_0 \ {\rm modulo}\ 2}
 where $p_1, \dots, p_{n-1}$ are positive and $p_n$ is negative
 (``chirality rule").  The $n$ points where the operators are inserted
 can be connected by a set of oriented lines that form a tree, with
 the condition that the tree intersects each loop at most once.  An
 example for $n=4$ is given in Fig.  10, where the fourth point is at
 infinity.  To each line of the tree we associate a charge.  The
 charges associated with the external lines are those introduced by
 the order operators.  The charges associated with the internal lines
 are determined by the charge neutrality of the vertices of the tree.
 In terms of momenta the neutrality condition for a vertex with $k$
 lines is written as $\sum _l (p_l -p_0) +2p_0=0 $ modulo 2.  Then the
 loops crossed by a line $l$ of the tree change from $2\cos(\pi p_0)$
 to $2\cos(\pi p_l)$, where $p_l = p_0-2q_l$ is the momentum
 associated with this line.  The above argument hold unchanged if
 instead of point-like insertions one considers finite boundaries.

%%%%%%%%%%%%%%%%%%%%%%%%%%%%%%%%%
\epsfxsize=200pt
\vskip 20pt \hskip 55pt \epsfbox{ 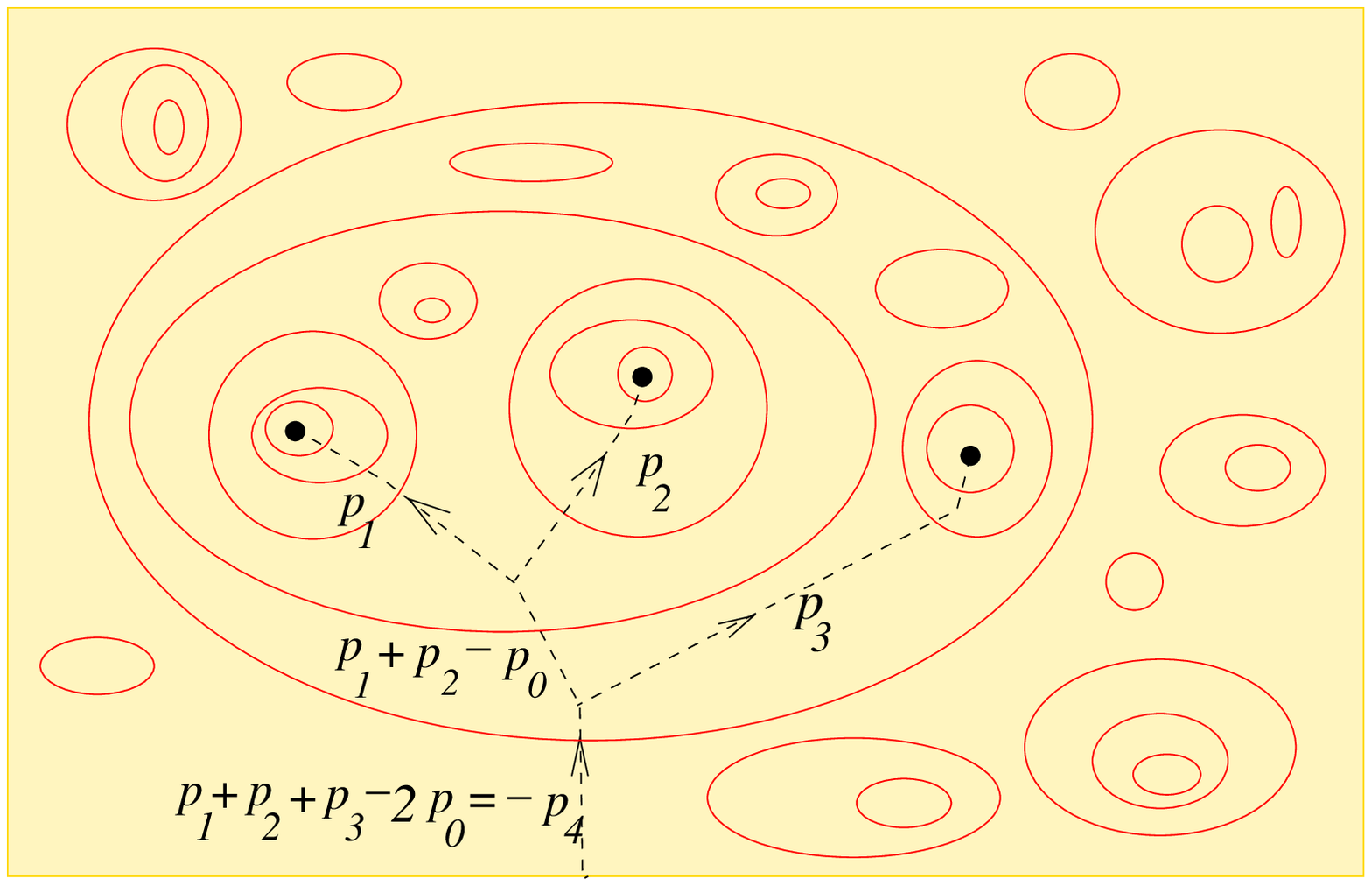 } \vskip 5pt
%%%%%%%%%%%%%%%%%%%%%%%%%%%%%%%%%%

\centerline{\ninepoint Fig.  10: A loop configuration in presence of
four operator insertions.  }

\vskip 10pt

\subsec{Correlation functions of order operators in the \SRSOS model}

\noindent Unlike the SOS model with $p_0\ne 0$, the \SRSOS model has a
well defined {\it partition function} on the sphere without
boundaries.  Summing over all height configurations yields a factor
$2\cos \pi p_0$ per loop.  Thus the partition function of the \SRSOS
model on the sphere coincides with the one for the $O(n)$ model with
$n = 2\cos(\pi p_0)$.

The {order operators} are introduced as insertions of the wave
functions
  \eqn\orderpD{\psi_{p}(x) = { \sin(\pi p x) \over \sin(\pi p_0 x)} ,
  \qquad p>0.  }
 If we restrict the spectrum of momenta to
\eqn\spcp{ p= mp_0, \quad m=1,2,...  } then the functions \orderpD\
generate  the $su(2)$  character ring.  

\bigskip \noindent {\it $\bullet$\ One-point function}

\noindent The only non-trivial one-point function is that of the
identity operator, $p=p_0$.  One can perform the sum of the heights by
repeatedly applying the relation \SAx\ starting with the domains that
do not contain loops.  As a result, the expectation value $\langle
\psi_{p_0} \rangle$ is equal to the partition function of the loop gas
on a sphere with one marked point, with weight $2\cos(\pi p_0)$ for
each loop.

\bigskip
\noindent
{\it $\bullet$\   Two-point  function}

\noindent The insertion of two operators \orderpD\ at two points
changes the weights of the non-contractible loops to $2\cos\pi p$.
The weights of these loops follow from the identity \SAx\ with $p_0$
replaced by $p$.  This two-point function is defined for any real
value of $p$.

\bigskip \noindent{\it $\bullet$\ $n$-point functions\ \ ($n\ge 3$)}

\noindent The $n$-point functions in the \SRSOS model are well defined
when the momenta belong to the spectrum \spcp,
\eqn\nptff{ G_{m_1,...,m_n}= \< \psi_{p_1}\psi_{p_2 } \dots
\psi_{p_n}\>, \qquad p_i = m_i p_0.  } One can again describe the
weights by drawing a tree connecting the $n$ points and such that each
loop is intersected at most once.  The momenta $m_1,...,m_n$ in
\nptff\ are associated with the external lines of the tree.  Unlike
the SOS case, here the momenta associated with the internal branches
are not determined uniquely.  When calculating the correlation
function with $n>3$, a sum over these momenta is to be performed.  The
weights in the momentum space are associated with the vertices of the
tree.  If $m_{l_1},...,m_{l_k}$ are the momenta associated with a
$k$-vertex, then the weight of this vertex is given by the $su(2)$
multiplicity $N_{m_{l_1},...,m_{l_k}}$, which is the Fourier image of
the weight \weightD\ with $n=k$.

It follows from the explicit form of the $su(2)$ multiplicities that if the
$n$ external momenta satisfy the neutrality condition
\eqn\saturm{ m_1+\dots +m_{n-1} -m_n = (n-2), } then the internal
momenta are uniquely determined and the $n$-point correlation function
\nptff\ coincides with the $n$-point function in the SOS model with
$n-1$ positive momenta, $p_i = m_i p_0\ (1\le k\le n-1)$ and one
negative momentum, $ p_n = -m_n p_0$.  Thus the $n$-point functions in
the \SRSOS model satisfying \saturm\ are identical to the
corresponding amplitudes in the SOS model that satisfy the ``chirality
rule''.

In the general case eqn.  \saturm\ is replaced by \eqn\saturm{
m_1+\dots +m_{n-1} -m_n = (n-2) + 2k,\qquad k\in \IZ_+. } The
contribution of given loop configuration to the $n$-point function can
be written as a sum of SOS $n$-point functions with $k$ extra
operators $\hat \psi_{-p_0}(x)= e^{i \pi p_0 x}$, associated with the
domains containing the vertices of the tree.

         \bigskip

\appendix{B}{Expression for  the  operators $a^{\dagger}_{k,x}$
in terms of the collective field}

\noindent We will use the expression \bpayy\ of the creation operators
$a_{n,x}^{\dagger}$ as contour integrals of the effective potential
$\G_x(z)$.  The operators $a_{n,x}^{\dagger}$ is proportional to the
coefficients in the expansion of the effective potential in the
half-integer powers of $z+M$.  In the $\t$ parametrization 
$$
\sqrt{1+\frac{z}{ M}}= \sqrt{2} \cosh\frac{\t} {2}\,.
$$
Instead of expanding $\G_x(z)$ at $z=-M$ we will evaluate the
integrals \alphxa\ by deforming the contour $\CC_-$ to the contour
$\CC_+$ going around the second cut of $\G_x(z)$ along the interval
$[M,\infty]$ as in Fig. 5b.  Deforming the contour $\CC_-$ to $\CC_+$ (with
clockwise orientation) we get, up to an infinite constant,
\eqn\bCy{ \eqalign{ {1\over \sqrt{2} \gh} \, \sum_{n=0}^{\infty} {u^n
\over n!} (a^{\dagger}_{n, x} -\d_{k,1} {S_x\over \gh}) &=\oint\limits_{\CC_-}
{dz\over 2\pi i}\ { \p _z \G_{x}(z)\over \sqrt{ 1+{z\over M} -2u}} =
\oint\limits_{\CC_+} {dz\over 2\pi i}\ { \p _z \G_{x}(z)\over \sqrt{
1+{z\over M} -2u}} \cr }}

The denominator in the last integral is regular  and we express 
the discontinuity of the  field
$\G_x(\t(z))$ in terms of $\Phi(\t(z))$. To evaluate the
integral we go to the $\t$ parametrization
\eqn\bptt{ {1\over \sqrt{2}\gh} \, \sum_{n=0}^{\infty} {u^n \over
n!}(a^{\dagger}_{n, x} - \d_{k,1}{S_x\over \gh} ) = 2 \, \int_0^{\infty}{d\tau
\over\pi}
    { \pt \sin \pi  x \p_\t \,  \cos \pi   \p_\t  \, \Phi^{-}(\tau) \over 
      \sqrt{1+\cosh\tau-2u}} %|0_\tw\rangle \,.
}
In this step we have retained the negative frequency part of the
field, i.e., \bptt\ is understood to hold applied on the right vacuum
$|0_\infty\rangle$.  The projections \bptt\ are seemingly non-local,
but they are actually given by polynomials of the derivatives of the
holomorphic field $\vp(\t)$ at $\t=\pm i \pi x$.  To see that we
evaluate the integral for each Fourier mode in the expansion \mdexp\
of $\vp^{-}(\t)$ using the integral formula for the Legendre function
of first kind \Bateman:
\eqn\legg{ \int_0^{\infty} {d\tau\over\pi} { \cosh \nu \tau\over
\sqrt{{\cosh\tau + 1 } -2 u} }= {P_{-\half +\nu}(1-2u)\over \sqrt{2}
\cos\pi \nu },}
\eqn\Piip{ P_{-\half +\nu}(1-2u) = \sum _{k=0}^{\infty} { u^k \over k!
} {(\hf+\nu)_k(\hf-\nu)_k\over k!  }, \ k=1,2,...}
The integral over $\nu$ amounts in replacing $\nu \to \p_\t$ and
setting $\t=0$, i.e., we get
\eqn\TauD{ a^{\dagger}_{n, x} -\d_{k,1} {S_x\over\gh}
=2\, { ({\hf} + \pt)_k
({\hf} - \pt)_k \over k!} \p_\t { \sin\pi x\pt } \, \Phi^-(\t)
\Big|_{\tau=0} } 
or, transforming to the momentum space and using the shorthand
notation \defFn\ for the differential operator,
\eqn\akdpa{\eqalign{ (a^{\dagger} _{k}(p)-\d_{k,1}\gh^{-1}
\delta^{(2)}(p-p_0))|0_\infty \rangle & = 
%2
\gh \,F_k(\p_\t)\, \p_\t
\vp^{-}(p,\t)\Big|_{\t=0} |0_\tw\rangle\cr &= i  \, F_k(\p_\t)\,
\sinh\t \, \p\G^{-}(\t+i\pi, p )\Big|_{\t=0} |0_\infty \rangle\,.  }}
 
On the other hand assuming that we can identify $|0_\infty\rangle$
with the right twisted vacuum $|0_\tw\rangle$, we can replace in the
r.h.s. $\p\G^{-}(\t+i\pi)$ with the twisted negative mode part.  Using
that the decomposition \alphxa\ takes the form
\eqn\alpshift{ \p \G_x^\dagger(\t +i\pi ) =  i\, {1\over M}
\sum_{n\ge 0} (a^{\dagger}_{n,x} - \delta_{n,1}{S_x\over \gh} ) \
{(-2)^{n-1}
(\sinh\frac{\t}{2})^{2n-1}\over (2n-1)!!} }
and plugging \alpshift\ in the identity \akdpa\ we see that it implies
that the operators $F_k$ act as projectors for $\t=0$, namely,
\eqn\difrela{ F_k(\p_\t) \, \sinh \t \ {2^{n-1}(-1)^n
(\sinh\frac{\t}{2})^{2n-1}\over (2n-1)!!} |_{\t=0}=\delta_{k, n},
\qquad n \ge 0 }
This property is also checked independently to hold true, which in
turn ensures the identification of the right vacua above.

\appendix{C}{The $D_\infty$  model}

The target space of the $D_\infty$ model is labeled by the positive
integers $x\in \IZ_+$ and the two extremities of the `fork' $x=0$ and
$x=\bar 0$ (Fig.  11).  The adjacency matrix is given by
\eqn\adjmd{\eqalign{ A^{x x'} &=\delta_{x, x' +1}+\delta_{x, x'-1},
\qquad x,x' \in {\Bbb N},\cr A_{0x} &= A_{\bar 0 x} =\delta_{1,x},
\qquad \qquad\ \ x \in {\Bbb N}.  }}

%%%%%%%%%%%%%%%%%%%%%%%%%%%%%%%%%
\epsfxsize=180pt
\vskip 20pt
\centerline{
\epsfbox{ 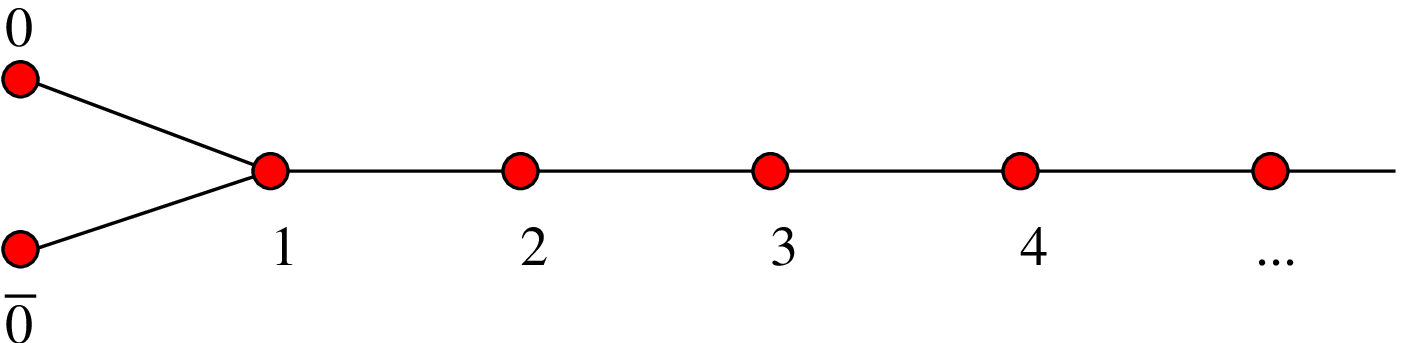   }
}
\vskip 5pt
%%%%%%%%%%%%%%%%%%%%%%%%%%%%%%%%%%

\centerline{\ninepoint
Fig. 11:  The target space of the $D_\infty$  model      }

\vskip  10pt
\noindent

Its eigenvectors $S^{(p)}_x$, $0\le p\le \pi$, are given by
\eqn\eignD{ S^{(p)}_x = \cases{ \cos( \pi px ) &if $x\in \IZ_+$;\cr \ \
\ \ \hf & if $x=0$ or $x=\bar 0$.}, \qquad 0\le p\le \pi.} The zeroth
eigenvalue $A(p)=0$ or $p=\hf$ is twice degenerated.  The second
eigenvector, which we denote by $\bar S^{(1/2)}$, has components
\eqn\degS{ \bar S^{(1/2)}_x = \cases{ \ \hf & if $x=0$;\cr -\hf & if
$x=\bar 0$;\cr \ 0& if $x\in\IZ_+$.  } } For $p_0= {1\over 2n+2}$ the
$D_\infty$ target space can be restricted to the $D_n$ Dynkin graph.

\listrefs

  \bye

 \appendix{D}{The target space CFT  for  pure  gravity}
 
 \noindent The simplest example of a conformal invariant field with
 non-conformal invariant expectation value is the operator solution
 for the pure 2d quantum gravity, which describes the possible
 critical behaviors of the one-cut solutions of the hermitian
 one-matrix model.  The partition function of pure gravity is defined
 by an appropriate scaling limit of the one-matrix integral.  The
 scaling limit consists in blowing up the vicinity of the right edge
 of the eigenvalue distribution, whose support after taking the limit
 extends to the whole negative semi-axis.  In the series of
 multicritical models including pure gravity the matter field is
 described by the $(2, 2m-1)$ minimal CFT. The simplest point in this
 series is not pure gravity $(2,3)$ but rather the Òpure topologicalÓ
 or ``gaussian'' theory $(2,1)$ with matter central charge $c=-2$.
 Nevertheless we will call all these theories ``pure gravity''.  The
 spectrum of operators in this theory is described by the flows of the
 KdV hierarchy \refs{\DVV, \FKN, \WittenKon} .  An important property
 of the theory is that there is a conserved charge (Òghost numberÓ)
 with a distinct background charge for each genus, so a specific
 correlation function can be non-zero for at most one genus.  It has
 been shown that the partition function of pure gravity is the only
 solution of the Virasoro constraints and that the ``string equation''
 for pure gravity is equivalent to the lowest Virasoro constraint,
 given that the partition function is a $\t$-function of the KdV
 hierarchy \refs{\DVV, \FKN}.

The target space CFT for pure gravity is defined on a Riemann surface
covering twice the complex plane and having a branched cut along the
negative real axis.  It is constructed out of the modes of the
$Z_2$-twisted gaussian field $\phi(e^{2\pi i } z ) = - \phi(z)$ with
mode expansion \eqn\cltt{\p \phi(z) = \sum_{r \in \IZ+\hf} \alpha_r \,
z^{-r-1}} where
\eqn\comRs{[\alpha_r, \alpha_s] = r\delta_{r+s, 0}.}
The twisted Fock vacuum $|0 \rangle $ and its conjugated $\langle 0 |
$ are defined as
 \eqn\Fcvac{\alpha_r |0\rangle= \langle 0|\alpha_{-r}=0\quad (r\ge
 1/2).}
We assume that the bosonic field has a non-zero classical value
$\phi_c$, which is again expanded in the half-integer powers of $z$
 \eqn\cctt{ \p \phi_c (z)= \sum_{r\in 1/2+ \IZ} r\, t_{2r} \, z^{r-1}
 .  } Here $\k$ is the string coupling constant and the coefficients
 $t_{2k+1}, \ k\in \IZ_+,$ are the coupling constants for the scaling
 operators in pure gravity.

 Our aim is to construct a right vacuum state which is conformal
 invariant.  The twisted vacuum itself does not satisfy all the
 Virasoro constraints, in particular it does not satisfy $L_{-1}$.
 Therefore we must dress the bare vacuum $|0\rangle$ by the modes of
 the bosonic field in such a way that all Virasoro constraints
 including the generator of translations $L_{-1}$ are satisfied.  If
 we denote by $\O_t$ the dressing operator, then it should satisfy
 \eqn\confS{ L_n \, \O_t |0 \rangle =0, \quad n \ge -1 } where the
 Virasoro operators \eqn\lmOm{ L_n = {\hf} \sum_{r+r'=n} :\alpha_r
 \alpha_{r'}: + \sum_{r-r'=n} \a_r \, r'\, t_{2r'}+ \frac{1}{8}
 \delta_{n, -1} \, t_{1}^2 \, + \frac{1}{ 16 }\, \delta_{n,0} } are
 associated with the stress energy tensor
  \eqn\vsciRR{ T(z) =\sum_n L_n \, z^{-n-2} ={ :\p[\phi (z) +\phi_c
  (z) ] ^2: \over 2} + {1\over 16 z^2 } .  } It is known that for each
  non-trivial classical background of the form \cctt\ the problem has
  a unique solution \refs{\DVV}.  The corresponding partition function
  is given by the expectation value
$$
Z_t=\< 0| \O_t|0\>.
$$
The partition function $Z_t$ satisfies the differential Virasoro
constraints obtained by replacing in \lmOm\ $$ \a_{-r} = r \, t_{2r},
\quad \a_{r} = {\p\over \p t_{2r}} \qquad (r\ge {\hf}).  $$

{\ninepoint The derivatives of the susceptibility $u=- \p^2 \log Z/\p
t_1^2$ in the couplings $t_{2k+1}$ are described by the KdV flows $$
{\p _{2k+1}} u ={\p _1} R_{2k+1} [u]\ \ \Leftrightarrow \ \
R_{2k+1}[u]= - \p_{2k+1} \p_1 \log Z $$
with $ R_1=u, R_3= {1\over 2} ( 3
%\k^2
 u^2 + \p_1^2 u)$ etc.  The $k$th critical point of pure gravity is
 described by the ``string equation'' $R_k[u]= t_1$.  The string
 equation for a generic perturbation is $$ \sum_{k=0}^{\infty} (2k+3)
 t_{2k+3} \, R_{2k+1}[u]+t_1=0.
$$
In the literature the operators generating the KdV flows are usually
denoted by $\s_n, n\ge 0$, see {\it e.g.}, \DVV. They are related to
the annihilation operators in the mode expansion \cltt\ of the twisted
field by
\eqn\defs{ \s_k = \sqrt{2}\, \a_{k +{1\over 2}} \qquad (k=0,1,2,
\dots)\, .  } }

\def\hty{{ \hat t}}

It is actually sufficient to construct the operator $\O_t$ for some
reference background $\hat \phi_c$ characterized by the couplings
$\hat t= \{ \hat t_{2k+1}\}_{k\ge 1}$.  Then the generic background $
t$ can be obtained by multiplying with an evolution operator $
e^{H[t-\hat t]} $:
      $$ \O_{t } = e^{H[t-\hat t]} \, \O_{\hat t }$$
      where the Hamiltonian $H$ is given by with \eqn\Hamyy{ H[t-\hty]
      =\sum _{r\ge 1/2} \, (t_{2r} - \hat t_{2r}) \, \a_r .  } The
      partition function of topological gravity can be written as the
      expectation value
 \eqn\pftg{ Z_t = \langle 0 | e^{H[t-\hat t]} \, \O_{\hat t
 }|0\rangle.  } The simplest reference background
\eqn\gaUs{\p\hat \phi_c(z) = - {1\over \kappa} \sqrt{z\over 2 },
\qquad \Leftrightarrow \qquad\hat t_{2k+1 } =- {\sqrt{2}\over {3} \k}
\ \delta_{k, 1} ,}
 corresponds to the gaussian, or topological, critical point of 2D
 gravity with matter central charge $c=-2$.  It describes the scaling
 limit of the gaussian matrix model.

The perturbative solution of the Virasoro conditions \lmOm\ with
couplings $t=\hat t$ can be obtained as a formal series in the
creation operators $\a_{-r}$ of the type
   \eqn\defstar{ \O_{t}(0)=
 \exp\( \sum_{n \ge 0} {1 \over n!} \sum_{k_1, \ldots ,k_n\ge 0} \
 w_{k_1 , \ldots k_n } \ \a_{-k_1-{1\over 2}} \ldots \a_{-k_n-{1\over
 2}}\) \, .  } The coefficients in \defstar\ are the correlation
 functions for the scaling operators \defs\
     generating the KdV flows.   
     : \eqn\wkkk{ w_{k_1 ,..., k_n}[t]=
      \< \s_{k_1}\cdots \s_{k_n}\>_{\hat t} .  }

 The scaling characterizing the gaussian point \gaUs\ is such that
 only finite number of correlation functions are non-zero for given
 genus $g$.  Given the quantum numbers $k_1,...,k_n$ in \wkkk, the
 genus of the non-vanishing amplitudes is determined by \WittenKon
 \eqn\gnus{\sum_{j=1}^n r_j ={3\over 2} (2g-2+n) \ \ \
 \Leftrightarrow\ \ \ \sum_{j=1}^n k_j = 3(g -1)+n \qquad (k_j =
 2r_j-1).} This should be understood as follows in the language of 2D
 quantum gravity: in the topological gravity the cosmological constant
 here coincides with the boundary cosmological constant $z$.  The
 string coupling constant $g_s$ has dimension of $z^{-(1- {1\over 2}
 \g_{\rm str})}$ with $ \g_{\rm str}= -1$.  The operator $\a_r$ has
 dimension of $z^{r}$.

  In order to simplify the numerical factors, we will rescale the
  operator amplitudes $\a_r$, following \WittenKon, as follows:
    \eqn\alphp{\eqalign{ \a_{-n-1/2}&=-\frac{1}{ \sqrt{2}} { 1\over
    (2n-1)!!  } \, {\t}_n^{\dagger} \cr \a_{n+1/2}&= -\frac{1}{
    \sqrt{2}} (2n+1)!!  \, \t_n } \qquad (n\ge 0).  } The rescaled
    creation and annihilation operators satisfy the canonical
    commutation relations
$$
[ \t_n, \t_k^{\dagger} ]= \delta_{n,k}.
$$
The lowest correlation functions
\eqn\coefF{ w_{k_1...k_n} [\hat t] = \(-\frac{1}{{\sqrt{2}}}\)^n
\prod_{i=1}^n (2k_i+1)!!\ \langle \t_{k_1 } ...  \t_{k_n}\rangle_{_g }
, } where by $\< \ \ \>_g$ we denote the genus $g$ expectation value,
are given by \IZk \eqn\tauSS{\eqalign{ &\langle \tau_ 1^n \rangle_{_1
}={(n-1)!\over 24}, \ \langle \tau_ 0^n\tau_{n+1} \rangle_{_1
}={1\over 24}, \ \langle\tau_ 0 \tau_ 1 \tau_ 2 \rangle_{_1 }={1\over
12}, \cr &\langle \tau_ 2 ^3 \rangle_{_2 } = {7\over 240}, \ \langle
\tau_ 2 \tau_3 \rangle_{_2 }= {29\over 5760},\ \langle \tau_
4\rangle_{_2 } = {1\over 1152}.\cr}}
The genus zero ($g=0$) correlation functions are simply the
multinomial coefficients \refs{\DVV} \eqn\ssts{ \langle \t _{m_1}...\t
_{m_n}\rangle _0 = {(m_1+...+m_n)!  \over m_1!...m_n!} , \ \
m_1+...+m_n= n-3.}

 The scaling law \gnus\ means that if $t_1=0$, then the genus $g$ free
 energy is given by a finite number of terms.  This is the case when
 the branch point of the classical background $\p\phi_c$ is at the
 origin.  In general the branch point can be at any point $a$ in the
 complex plane.  We can achieve that $t_1=0$ by shifting $z\to z-a$ in
 the expansion \cctt.  The coefficients in the expansion at $z=a$ are
 called {\it moments} of the classical background.  Of course, the
 moments are not independent coordinates because the expansion point
 depends on the background.  If $t_1=0$, then $\mu_n \sim t_{2n+1}$.
 The moments are defined by the generating function
  \eqn\moMts{ 
   \sum_{n\ge 0} {\mu_n u^n\over n!}
   =- \oint {dz\over 2\pi i} 
{\p \phi_c(z)\over \sqrt{{1\over 2} (z-a) - u}}, \qquad n\ge 0.
}
 The position of the branch point is determined by the condition 
 $\mu_0=0$.   Then the  genus $g$ free energy   for {\it any}  
  background of the form  \cctt \ is given by the finite sum \IZk :

\eqn\Fmom{\eqalign{ F_1&= - \frac{1}{24} \log \mu_1 , \quad F_2= {29
\mu_2\mu_3\over 5760 \mu_1^4} + { \mu_4 \over 1152 \mu_1^3}
++{7\mu_2^3\over 1440 \mu_1^4},\cr F_g &= \sum_{n\ge 0}
\frac{\mu_1^{2-n-2g}}{ n!} \sum_{^{ k_1, ..., k_n \ge 2 }_{
k_1+...+k_n = n+3g-3}} \< \t_{k_1}\cdots \t_{k_n}\> _g\
\mu_{k_1}\cdots \mu_{k_n} .} } 
The moment $\m_n$ is a linear combination of the couplings $t_{2k+1},
\ k\ge n$, with coefficients obtained by expanding the integrand in
\moMts\ in half-integer powers of $z$.  The possibility of such a
representation of the free energy in terms of a finite number of
moduli (moments) characterizing the corresponding Riemann surface is a
general property of any matrix model background even before taking the
scaling limit \ackm .

\bye